%% file: main.tex
\documentclass[apjpt4]{aastex}

\newcommand{\civfull}{C~{\sc iv}~$\lambda$1549}
\newcommand{\civ}{C~{\sc iv}}
\newcommand{\feka}{Fe~K$\alpha$}
\newcommand{\luv}{$l_\nu$(2500~\AA)}
\newcommand{\lx}{$l_\nu$(2~keV)}
\newcommand{\aox}{$\alpha_{\rm ox}$}
\newcommand{\luvunit}{erg~s$^{-1}$~Hz$^{-1}$}

\newcommand{\siiv}{Si~{\sc iv}~$\lambda$1396}
\newcommand{\mgii}{Mg~{\sc ii}~$\lambda2798$}
\newcommand{\lya}{Ly$\alpha$}

\newcommand{\ciii}{C~{\sc iii}]~$\lambda1908$}
\newcommand{\oii}{[O~{\sc ii}]~$\lambda3727$}
\newcommand{\nev}{[Ne~{\sc v}]~$\lambda3426$}
\newcommand{\dew}{$\Delta\log{\mbox{EW}}$}
\newcommand{\xray}{\hbox{X-ray}}
\usepackage{amsmath}

\shorttitle{\civ\ and \feka\ BEff} \shortauthors{Wu et al.}

\begin{document}
\title{Probing the Origins of the \civ\ and \feka\ Baldwin Effects}

\author{Jian Wu\altaffilmark{1}, Daniel~E.~Vanden Berk\altaffilmark{1,2}, W.~N.~Brandt\altaffilmark{1}, Donald~P.~Schneider\altaffilmark{1},
Robert~R.~Gibson\altaffilmark{1,3}, \\
and\\
Jianfeng~Wu\altaffilmark{1}}
\email{jwu@astro.psu.edu}
\altaffiltext{1}{Department of Astronomy \& Astrophysics, the Pennsylvania
State University, 525 Davey Lab, University Park, PA, 16802, USA}
\altaffiltext{2}{Department of Physics, Saint Vincent College, 300 Fraser-Purchase Road, Latrobe, PA, 15650, USA}
\altaffiltext{3}{Department of Astronomy, University of Washington, Box 351580, Seattle, WA, 98195, USA}

\begin{abstract}
We use UV/optical and X-ray observations of 272 radio-quiet Type 1 
AGNs and quasars to investigate the \civ\ Baldwin Effect (BEff). 
The UV/optical spectra
are drawn from the \emph{Hubble Space Telescope}, \emph{International
Ultraviolet Explorer} and Sloan Digital Sky Survey archives. The
X-ray spectra are from the \emph{Chandra} and \emph{XMM-Newton}
archives. We apply correlation and partial-correlation analyses to
the equivalent widths, continuum monochromatic luminosities, and
\aox, which characterizes the relative X-ray to UV brightness. 
The equivalent width of the \civfull\ emission line is correlated with both
\aox\ and luminosity. We find that by
regressing \luv\ with EW(\civ) and \aox, we can obtain tigher correlations
than by regressing \luv\ with only EW(\civ). Both correlation
and regression analyses imply that \luv\ is not the only factor
controlling the changes of EW(\civ); \aox\ (or, equivalently, the
soft \xray\ emission) plays a fundamental role in the formation and 
variation of \civ. Variability contributes at least 60\% of the scatter of 
the EW(\civ)-\luv\ relation and at least 75\% of the scatter of the 
of the EW(\civ)-\aox\ relation.

In our sample, narrow \feka\ 6.4 keV emission lines are detected in 50
objects. Although narrow \feka\ exhibits a BEff similar to
that of \civ, its equivalent width has almost no dependence on
either \aox\ or EW(\civ). This suggests that the majority of narrow \feka\ 
emission is unlikely to
be produced in the broad emission-line region. We do find suggestive
correlations between the emission-line luminosities
of \civ\ and \feka, which could be potentially used to estimate the
detectability of the \feka\ line of quasars from rest-frame UV 
spectroscopic observations.
\end{abstract}



\clearpage
\input{sec1}

\input{sec2}

\input{sec3}

\input{sec4}

\input{sec5}

\input{sec6}
\acknowledgments 
\par
We thank Jane Charlton for
providing a number of \emph{HST}/FOS spectra of the core sample
AGNs, Eric Feigelson for useful suggestions and advice
on statistics, Ohad Shemmer and Dennis Just for discussions on
linear regression, and Lanyu Mi for help with the statistical
computations. 

This work was partially supported by NSF grant AST-0607634 and NASA LTSA grant
NAG5-13035.

Funding for the SDSS and SDSS-II has been provided by the Alfred P. Sloan 
Foundation, the Participating Institutions, the National Science Foundation,
the U.S. Department of Energy, the National Aeronautics and Space 
Administration, the Japanese Monbukagakusho, the Max Planck Society, and the 
Higher Education Funding Council for England. The SDSS website is 
\url{http://www.sdss.org/}.

\input{bib} 
\input{figure}
\input{table} 

\end{document}

%% file: sec1.tex
\section{Introduction}
\label{introduction} One of the important properties of AGNs
is the relation between the emission-line strength,
characterized by the equivalent width (EW), and continuum
luminosity, because it reveals that the regions emitting these
two spectral components are associated. 
\citet{bal77a} found that the EW of \civfull\ (\civ)
is inversely correlated with the quasar monochromatic luminosity
at 1450~\AA, $l_\lambda$(1450~\AA), namely, 
$\log{\mbox{EW(C~{\sc iv})}}=k\log{l_\nu(1450~{\mbox{\AA}})}+b$. 
\citet{car78} referred to this
trend as the ``Baldwin Effect" (BEff), a designation now widely used
to describe line strength-luminosity relations. \citet{bal77a}
identified this relation using only 20 quasi-stellar objects with
$29.8\lesssim\log{l_\nu(\mbox{1450\ \AA})}\lesssim32.0$ and
$1.24<z<3.53$. Subsequent UV/optical surveys have enabled 
investigation of this relation with wider luminosity and redshift ranges (e.g.,
Kinney, Rivolo \& Koratkar 1990; Zamorani
et al. 1992)\nocite{kin90,zam92}. It has been found that the BEff
exists for not only \civ\ but many other broad emission lines such
as \lya, \ciii, \siiv, \mgii\ \citep{die02,van08}, UV iron
emission lines \citep{gre01}, and even forbidden lines
such as \oii\ and \nev\ \citep{cro02}. Applying a spectral-composite technique
\citep{van01} to the Sloan Digital Sky Survey (SDSS; York et al.\
2000) \nocite{yor00} Data Release Three (DR3) quasar catalog
\citep{sch05}, \citet{van08} found that the BEff evolves with
redshift, which is a source of scatter in this relation for a
sample with a wide range of redshift.

The X-ray BEff (or \feka\ BEff), in which the EW of the narrow 
\feka\ line at 6.4~keV (hereafter abbreviated to \feka) is
anti-correlated with X-ray luminosity, \lx, was discovered in the
early 1990s from observations by the X-ray observatory
\emph{Ginga} \citep{iwa93}. This relation has been subsequently
confirmed using data from \emph{ASCA} \citep{tan94,nan97} and from
\emph{Chandra} and XMM-\emph{Newton} \citep{pag04a,zho05,jiang06,bia07}.
Possible sites of origin for narrow \feka\ emission include the broad 
emission-line region (BELR), the outskirts of the accretion disk, and the
molecular torus (e.g., Weaver et al. 1992, Antonucci 1993; Krolik, Madau \& 
Zycki 1994).
\nocite{wea92,ant93,kro94}

Although it is well accepted that the BEff exists for many
UV/optical emission lines (e.g., Osmer \& Shields\ 1999; Shields 
2007)\nocite{osm99,shi07}, there is currently
no theoretical model that provides a compelling and complete explanation
of this well-known phenomenon. Several physical explanations have
been proposed to account for the UV/optical BEff.

One promising explanation is that the continuum
shape may be luminosity dependent. In this model, the UV/optical 
BEff is due to the
softening of the spectral energy distribution (SED) at high
luminosity, which lowers the ion populations having high ionization
potentials \citep{net92,kor98}. It has been found, using
\emph{Einstein Observatory} data, that the quasar SED,
parameterized by \aox\footnote{Defined as $\alpha_{\rm ox}=\log\left[l_\nu(\mbox{2\ keV})/l_\nu(\mbox{2500\ \AA})\right]/\log\left[\nu(\mbox{2\ keV})/\nu(\mbox{2500\ \AA})\right]=0.3838\log{\left[l_\nu(2\mbox{\ keV})/l_\nu(\mbox{2500\ \AA})\right]}$. \aox\ is used to characterize the spectral hardness in the UV to 
X-ray band (e.g., Avni \& Tananbaum 1982, 1986;
Anderson \& Margon 1987; Wilkes et al. 1994; Vignali, Brandt \&
Schneider 2003; Strateva et al. 2005; Steffen et al.
2006; Just et al.  2007).\nocite{avn82,avn86,and87,wil94,vig03,str05,ste06,just07}} \citep{tan79}, depends on UV
luminosity (e.g., Zamorani et al. 1981)\nocite{zam81}. Later research using
radio-quiet (RQ) optically selected quasar samples from the SDSS
Early Data Release \citep{sto02}
confirmed and extended this result (e.g., Vignali, Brandt \& Schneider
2003)\nocite{vig03}. Using optically selected AGNs
\citet{str05} and \citet{steffen06} firmly established the
correlation of \aox\ with UV luminosity for these sources. The idea 
that the UV/optical BEff
is attributable to SED-driven ionization effects is supported 
both observationally
\citep{zhe93,gre96,gre98} and theoretically \citep{net92}. Recent work on 
a sample of non-Broad Absorption Line (BAL), radio-quiet, optically selected 
quasars indicates that
the EW of \civ\ depends both on UV and X-ray luminosity. The physics of the 
\civ\ BEff is apparently associated with both UV and X-ray emission 
(e.g., Gibson, Brandt \& Schneider, 2008, hereafter GBS08\nocite{gib08},
and references therein). 

Other proposed BEff drivers include the Eddington
ratio, $L/L_{\rm Edd}$ \citep{bas04,bac04,war04,zho05}, the
black-hole mass \citep{net92,wan99,shi07},
and the luminosity dependence of metallicity \citep{war04}.

In this paper we investigate the origin of the BEff for the \civ\
emission line in a sample of 272 Type 1 AGNs and quasars. Although
\civ\ is not the only UV/optical broad emission line that exhibits
a BEff, we selected it to study this phenomenon not only because it
is a representative and well-accepted BEff emission line, but also
because \civ\ resides in a relatively clean spectral region where
the local continuum can be well approximated as a single
power-law, with few blends with other emission lines (in particular the iron 
emission forest) and limited contamination from the AGN host galaxy.
These properties make it relatively straightforward to perform spectral fitting
and obtain accurate emission-line parameters for \civ.
We also use partial-correlation analysis (PCA) and
linear-regression regression analysis to investigate the correlations
between EW, monochromatic luminosity, and \aox\ for \civ\ and narrow \feka\
emission lines. 

Over the past three decades, there have been a large number of studies of the
BEff. Our work on the \civ/\feka\ lines combines the 
following important features (1) a wide range in redshift 
($0.009\lesssim z\lesssim4.720$)
and luminosity ($27.81\lesssim\log{l_\nu(2500~\mbox{\AA})}\lesssim33.04$) that
allows one to disentangle evolutionary vs. luminosity, so that we 
are not narrowing our study for quasars with a particular luminosity 
or at a certain redshift; (2) a relatively large 
sample size (272 objects); (3) the use of partial correlation analysis; and 
(4) a high X-ray detection rate ($\sim94\%$); and (5) estimates of the effects 
of observational errors and object variability.

We describe the sample selection in
\S~\ref{data} and the methods used to process the data in
\S~\ref{data_reduction}. In \S~\ref{drive}, we perform partial-correlation 
and linear-regression analyses to investigate the roles of \aox\ in the
\civ\ and \feka\ BEffs. In \S~\ref{relation}, we
probe the connections between \civ\ and \feka\ relationships in EWs, fluxes
and luminosities. We present our conclusions in \S~\ref{conclusions}.
Throughout this work, we adopt the following cosmology:
$\Omega_{\rm M}=0.3$, $\Omega_\Lambda=0.7$, $H_0=70$
km~s$^{-1}$~Mpc$^{-1}$.

%% file: sec2.tex
\section{Sample Construction}
\label{data} The quasar sample in our study is drawn from three
sources: 50 objects from \citet{jiang06}, which will be referred
to as ``Sample A"; 98 objects from GBS08\nocite{gib08}, which will
be referred to as ``Sample B"; and 124 objects from \citet{just07}, which
is referred to as ``Sample C". We define our ``combined sample" as
the combination of these three data sets.
\subsection{Sample A}
\citet{jiang06} compiled a dataset of 101 Type 1 AGNs with \feka\
observations from both the \emph{Chandra} and \emph{XMM-Newton}
archives. The detection fraction of the \feka\ line in their
combined sample is around 55\%. The redshifts of the AGNs range
from $0.003$ to $3.366$, but most of the objects (87\%) are low-redshift 
($z\lesssim0.4$) AGNs. The monochromatic luminosities,
\luv, range from $10^{26.0}$ to $10^{31.5}$ \luvunit. We chose this
dataset because it is the most complete \feka\ BEff sample with
high-quality data obtained from the most sensitive X-ray missions.
However, because the number of AGNs observed in the X-ray band is much smaller 
than the number observed optically and not all X-ray observed
AGNs present \feka\ emission lines, the \feka\ sample is
significantly limited in size.

We searched for UV/optical spectra of all 101 objects from the
archival databases for \emph{HST} and \emph{IUE}. We found 82 
spectra covering the wavelength region around the \civ\ emission
line (containing at least the 1500--1600 \AA\ band). If 
observations are available from
both \emph{HST} and \emph{IUE}, we selected the \emph{HST}
observations due to their generally higher signal-to-noise (S/N) ratio. 
For spectra with multiple observations using the same instrument, we
preferentially use spectra with higher S/N.

Next, we excluded all the radio-loud (RL) objects from the core sample,
because additional X-ray emission is produced by the radio jet (e.g., 
Brinkmann et al.\ 2000\nocite{bri00}) and changes the value of 
\aox\ as well as the slope of the \feka\ BEff.

We further excluded 3 objects for the following reasons:
\begin{itemize}

\item \object{\emph{MCG-06-30-15}}: The \feka\ profile of MCG-06-30-15 is
well fit using a broad disk line model (e.g., Tanaka et al. 
1995\nocite{tan95}). The narrow component is not well resolved or very weak.
In this work, we only study the narrow component of \feka, so this
object is excluded. 

\item \object{\emph{IC~4329a}}: The spectrum has low S/N, and the
\civ\ emission line is cannot be accurately measured (e.g., Crenshaw \& Kraemer 
2001\nocite{cre01}).

\item \object{\emph{PG~1407$+$265}}: This object was termed as an
``unusual" quasar \citep{mcd95} because it contains extremely weak
Ly$\alpha$ and \civ\ lines. Because of its peculiarity, we exclude
it from our sample.
\end{itemize}

We removed 5 objects with strong associated absorption lines of
\civfull:  
\object{PG~1411$+$442} \citep{wis04}, 
\object{NGC~4151} \citep{cre99}, \object{Ark~564} \citep{cre99}, 
\object{NGC~4051} \citep{col01}, and \object{PG~1114$+$445} \citep{sha07}. 
These features prohibit reconstruction of the unabsorbed \civfull\ 
emission-line profile. The X-ray absorption associated with the UV line 
absorption \citep{bra00} might also lead to an under-prediction of 
continuum flux at 2~keV and, therefore, an incorrect estimation of 
the intrinsic value of \aox.

Finally, we exclude 8 objects that are classified as Seyfert~1.5 
(Sy~1.5) and Sy~1.9 \footnote{Based on
NASA/IPAC Extragalactic Database: http://nedwww.ipac.caltech.edu/} 
\citep{ost81,ost89}. These objects are intermediate between Sy~1 and Sy~2 
galaxies and are often subjected to obscuration along the line 
of sight. Thus their UV, X-ray luminosities and \aox\ values are 
also potentially affected. 

The final version of Sample A consists of 50 AGNs. Among these objects,
34 are found in the \emph{HST} archive (FOS\footnote{Faint Object Spectrograph} 
or STIS\footnote{Space Telescope Imaging Spectrograph}), and 16 are
found in the \emph{IUE} archive (SWP\footnote{Short Wavelength Prime} 
or LWP\footnote{Long Wavelength Prime}). The \feka\ detection rate is 55\%, 
the same as in the entire \citet{jiang06} study. 
\subsection{Sample B}
Objects in Sample B were selected from 536 SDSS DR5 (Adelman-McCarthy et al.
2007\nocite{ade07}) quasars in
GBS08\nocite{gib08}. These quasars, taken from the DR5 Quasar Catalog 
\citep{sch07}, are at redshift $1.7\leq z\leq2.7$
and have been observed by \emph{Chandra} or XMM-\emph{Newton}. The
lower redshift limit ensures that all SDSS spectra in this
sample cover the \civ\ region;  the upper redshift limit
ensures that the rest-frame flux at 2500 \AA\ is covered so that \aox\
can be measured accurately.

We excluded the BAL 
quasars in Table~1 of GBS08\nocite{gib08}. BAL quasars can have
strong absorption features in the \civ\ spectral region that prohibit
accurate fitting of the continuum and emission-line profiles. In
addition, BALs are usually associated with relatively strong X-ray
absorption (e.g., Brandt, Laor \& Wills 2000; Gallagher et al. 
2006)\nocite{bra00,gal06}. We only retain objects with 
\emph{Chandra} observations 
with angular offsets $<10^\prime$ to avoid large X-ray flux
uncertainties caused by variations of the point spread function. 
This restriction reduces
our sample size to 149. In addition, we excluded RL
objects and strong associated absorption-line (AAL) objects.
The final version of Sample B consists of 98 objects with
\luv\ between $10^{30.53}$ and $10^{31.67}$ \luvunit. This is
a relatively narrow luminosity range; however, it does provide a significant
extension of Sample A since the latter is mostly composed of
low-luminosity and low-redshift AGNs.

\subsection{Sample C}
To examine the relations between
\aox, \luv, and \lx, \citet{just07} compiled a sample of 372
objects, including 26 from their core sample, 332 
from \citet{steffen06}, and 14 from \citet{she06}. BAL quasars, RL
quasars, and gravitationally lensed objects have already been
excluded from this sample. The gravitationally lensed quasars are
removed because their fluxes are strongly amplified and thus their
luminosities are uncertain.

Among the 372 objects, 38 are already in Sample A. For the rest of
the AGNs, we searched for existing spectra with \civ\ coverage
preferentially from SDSS, then the \emph{HST} and \emph{IUE} archives. 
Finally, we 
removed 5 AGNs whose spectra contain strong AALs. These restrictions leave
124 objects in Sample C, in which 91 objects are from the SDSS DR5 
quasar catalog, 13 from the \emph{HST} archive, and 20 from the 
\emph{IUE} archive.  The redshift of this sample ranges from
0.015 to 4.720 and \luv\ ranges from $10^{28.12}$ to $10^{32.32}$
\luvunit.

\subsection{Combined Sample} The combined sample (Table~\ref{tab-samsum}) 
consists of a total of 272 objects: 189 (69.5\%) have spectra from SDSS, 47
(17.3\%) from \emph{HST}, and 36 (13.2\%) from \emph{IUE}. The redshifts range 
from 0.009 to 4.720 (Fig.~\ref{fig-lz}). The gap between $z\sim0.5$
and $z\sim1.5$ is caused by instrumentation limitations. Because of the
wavelength coverage of the SDSS spectrographs, the redshifts of SDSS 
quasars having \civ\ coverage must be greater than $1.5$. Most 
intermediate-redshift AGNs ($0.5\lesssim z\lesssim1.5$) are too faint for
their UV/optical spectra to be taken by \emph{IUE} and \emph{HST}. Our 
sample exhibits a strong redshift-luminosity
correlation (Fig.~\ref{fig-lz}); we discuss this issue further 
in our analyses below. 

The \luv\ of this combined sample ranges between $10^{26.53}$ and
$10^{33.04}$ \luvunit, including Seyfert galaxies to the most-luminous
quasars in the Universe. The X-ray detection rate is $94.9\%$. The UV
properties of the combined sample as well as the UV and X-ray 
properties of Sample A are tabulated in Table~\ref{tab-sampleu} 
and Table~\ref{tab-sampleax}. The combined 
sample will be used to investigate the \civ\ BEff, while only 
Sample A will be used to study the \feka\ BEff.

%% file: sec3.tex
\section{Data Processing}
\label{data_reduction}
\subsection{Bad Pixel Removal and Reddening Correction}
To ensure that we use high-quality data to perform the continuum and 
emission-line fitting, we 
remove bad pixels in the SDSS spectra based on
the mask column contained in the SDSS quasar spectral files. 
We removed all the bad pixels in the spectra of SDSS objects in Sample 
C.\footnote{For data processing of Sample B, refer to 
GBS08\nocite{gib08}.} 
The excluded pixels cover less than 10\% of the total pixels 
for over 98\% of SDSS objects, and the maximum fraction of removed pixels 
for a single object is 15\%.

We perform Galactic reddening corrections to all the spectra using the
$E(B-V)$ dependent extinction curve of \citet{fit99}. Values of
$E(B-V)$ are calculated following \citet{sch98}.
\subsection{Spectral Fitting}
\subsubsection{Sample A}
For each UV/optical spectrum from \emph{HST} or \emph{IUE}, we fit the local
continuum in the vicinity of \civ\ (typically 1300--1700 \AA)
using a single power-law. We do not expect our measurements to be
significantly affected by host-galaxy components because 1)
our sample contains only Type~I AGNs 
and quasars in which emission from the nucleus dominates the
light from host galaxies in the UV band; and 2) \civ\ is in the UV
region in which the non-nuclear emission primarily arises from
massive stars such as O and B stars. Examination of the spectra does not
reveal any stellar absorption lines, indicating that the contribution from
starlight is negligible. We do not subtract the iron
emission forest as this component is usually not strong around the
\civ\ emission line (e.g., Shen et al. 2008)\nocite{she07}, and
the wavelength coverage of the \emph{HST} and \emph{IUE} spectra is
frequently too narrow (only a few hundred Angstroms) to fit this
component. The Balmer continuum ``small blue bump" only appears in
the wavelength range between 2000--4000~\AA, so its contribution
is negligible around the \civ\ region.

The emission-line spectrum is obtained after subtracting the
power-law continuum. We fit the \civ\ emission lines using two
Gaussian profiles. The model always produces visually acceptable fits. We
mask narrow absorption-line features appearing near the
emission-lines so as not to under-predict the emission line flux. The
equivalent widths, the emission-line luminosities under the
assumption of isotropy, and the continuum monochromatic
luminosities at 2500 \AA\ are calculated under our
adopted cosmology and are tabulated in Table~\ref{tab-sampleu}.

To first order, we use 2--10 keV luminosities
tabulated in Table 1 of \citet{jiang06} to estimate \lx\
under the assumption that the X-ray continuum
is a single power-law with photon index $\Gamma=2$ from 2 keV to
10 keV (e.g., Page et al. 2005; Shemmer et al. 2005; Vignali et al. 
2005)\nocite{pag05,she05,vig05}, so that 
\begin{equation}
\label{l2kevlx}
l_\nu({\rm 2\ keV})=\frac{L(\mbox{2--10\ keV})}{\nu_2\ln{5}}
\end{equation}
where $h\nu_2=2\mbox{ keV}$ and $h$ is Planck's constant. To obtain the
\feka\ emission-line flux, we further assume that the \feka\
emission line resembles a single Gaussian profile: $f_{\rm
l}(\nu)=A\cdot e^{-\left(\nu-\nu_0\right)^2/2\sigma^2}$. This
allows one to express the line flux $F_{\rm l}$ in terms of
$l_\nu$(2~keV) or $F$(2--10 keV), the \feka\ equivalent width
EW(\feka), and the central energy ($\epsilon_0=6.4$ keV) of
the emission line. Assuming the continuum flux $f_{\rm
c}=C\cdot\nu^{-1}$, we can derive
\begin{equation}
  \mbox{EW(Fe~K}\alpha)=\int_{0}^{+\infty}\frac{f_{\rm l}(\nu)}{f_{\rm
c}(\nu)}\,d\nu=\frac{A}{C}\left[\sigma^2e^{-\nu_0^2/2\sigma^2}
             +\nu_0\left(\sqrt{2\pi}\sigma-\int^{+\infty}_{\nu_0}e^{-x^2/2\sigma^2}\,dx\right)\right]\approx\frac{A}{C}\sqrt{2\pi}\nu_0\sigma.
\end{equation}
The \feka\ emission-line flux is
\begin{equation}
F_{\rm l}=\int_{0}^{+\infty}f_{\rm l}(\nu)\,d\nu
           =A\left(\sqrt{2\pi}\sigma-\int_{\nu_0}^{+\infty}e^{-x^2/2\sigma^2}\,dx\right)
           \approx A\sqrt{2\pi}\sigma.
\end{equation}
The approximations are valid because generally
$\nu_0\gg\sigma$ so $\nu_0^2/2\sigma^2\gg1$. For instance,
$\epsilon_0(\mbox{Fe~K}\alpha)=6.4$~keV while the width,
$\sigma$(\feka), is usually $\lesssim0.1$~keV; as a result,
$e^{-\nu_0^2/2\sigma^2}\approx0$. We can therefore calculate the \feka\
line flux by
\begin{equation}
\label{flfeka}
F_{\rm l}=\frac{\mbox{EW(Fe~K}\alpha)}{\epsilon_0}\frac{F(\mbox{2--10\ keV})}{\ln{5}}.
\end{equation}
The \feka\ EWs, emission-line luminosities, and 2 keV
monochromatic luminosities of Sample~A are tabulated in
Table~\ref{tab-sampleax}.
\subsubsection{Sample B}
For objects in Sample B, we directly adopt the fitting results from
GBS08\nocite{gib08}. In their paper, the SDSS spectral continua were
fit with polynomials, and the \civ\ emission lines were
fit with Voigt profiles. The different model in GBS08\nocite{gib08} from
our work used to fit the \civ\ spectral region will not cause 
significant differences; because the continuum around \civ\ 
is not contaminated with 
other emission/absorption lines, the polynomial fit will produce almost
the same result as the simple power-law fit. In addition, since both multiple
Gaussian and Voigt profiles produce acceptable fits to the emission line, they
will give nearly the same line flux. The X-ray spectral continua were fit 
using a broken power-law with the power-law break fixed at 2~keV in
the rest-frame in order to obtain \lx.
\subsubsection{Sample C}
We fit the UV/optical spectra from the SDSS using a
routine, described by \citet{van08}, developed for SDSS quasar
spectra. This routine fits the ``underlying continuum" using three
components simultaneously: a power-law, the iron emission forest,
and the small blue bump. We adopt the UV iron emission template
(1075--3090 \AA) from \citet{ves01} and the optical template
(3535--7534 \AA) from \citet{ver04}. We first make a preliminary
estimate of the power-law component by connecting two ``line-free" 
points in the spectra. This provides initial estimates of the power-law
parameters for the subsequent comprehensive processing in which the 
spectra are fit by considering all three components mentioned above. 
We evaluate the fitting quality by calculating $\chi^2$ values within some
``line-free" windows. Finally, the continuum fit is subtracted
from the original spectrum and the residuals are used to conduct
emission-line fitting. For Sample C, the \civ\ emission-lines are fit by
superpositions of two Gaussian profiles, which always yields
acceptable fits for the data. We then calculate the \civ\ EW, emission-line 
luminosity, and monochromatic luminosity at 
2500~\AA\ (Table~\ref{tab-sampleu}). We adopt
the values of \aox\ and 2~keV monochromatic luminosity from
\citet{just07}.

Examples of continuum and emission-line fits of the \emph{HST},
\emph{IUE} and SDSS spectra are presented in Fig.~\ref{fig-fit}.

\subsection{Error Analysis}
In order to estimate the uncertainties in the measured quantities for objects 
in Samples A and C, we ran Monte Carlo simulations assuming a model
with a perfect correlation between EW(\civ), $f(\mbox{\civ})$, and
$f_\nu(2500\mbox{~\AA})$. For each
spectrum, we add random noise to the original best fit to
produce artificial spectra. The random noise follows a Gaussian
distribution, and its amplitude is determined in one of two ways.
For a spectrum from the \emph{HST} or \emph{IUE} database, we
apply a low band pass filter, filtering out low-frequency signals via
fast Fourier transformation.\footnote{http://www.msi.umn.edu/software/idl/tutorial/idl-signal.html}
The residual signal is mostly noise. We then calculate the root-mean-square
(RMS) of the noise and take this value as
the random noise amplitude to be added onto the model spectrum.
For a spectrum from the SDSS database, we simply use the uncertainty level
associated with each pixel as the noise amplitude. For each
spectrum, we produce 100 artificial spectra and fit them in
exactly the same way as the observed spectrum. The
error bar of a spectral parameter is then calculated as the RMS 
of 100 fitting results.
The typical
error bar shown in each plot (e.g., Fig.~\ref{fig-civbeffl})
is the median of all the error bars of
points in that plot. 

When evaluating the uncertainties of monochromatic luminosities, e.g.,
\luv, we must consider the contribution from the error in the 
luminosity distances (important for the low-redshift objects). To 
validate this, we calculate the ratio of 
luminosity uncertainties without considering distance errors ($\delta_{f}$;
the error in the flux measurement) 
to luminosity uncertainties after considering distance errors 
($\delta_{f,d}$) for
objects in Sample A, and denote it as $\delta_{f}/\delta_{f,d}$, in 
which $f$ stands for flux and $d$ stands for distance. Assuming that
the square of total uncertainty can be expressed as the quadratic 
summation of the uncertainties of flux and distance individually, 
the square of the ratio, $(\delta_{f}/\delta_{f,d})^2$, is more 
relevant than the ratio itself. 
We find that there are 5 out of 50 objects (in Sample A) whose ratios 
are above 0.1;
three of them are even greater than 0.5. These objects have very low redshifts
but large redshift uncertainties ($\delta_z>0.001$). 
For consistency, it is necessary to include the distance into the gross 
uncertainty calculation for all the objects in our samples. 
These uncertainties are also
estimated using a Monte Carlo method. We adopt the most accurate
values of redshift and their uncertainties from the NASA/IPAC
Extragalactic Database.\footnote{http://nedwww.ipac.caltech.edu/}
These uncertainties are treated as noise amplitudes to be added to the
redshift values. The luminosity uncertainties are calculated using
the following error-propagation equation:
$$\delta L=4\pi d_{\rm L}\sqrt{\left(d_{\rm L}\delta f\right)^2+4f^2\left(\delta d_{\rm L}\right)^2}$$

We adopt a 20\% uncertainty for each X-ray continuum luminosity,
e.g., \lx; their measurement errors are not available in the 
literature. The relative uncertainty varies considerably depending 
on the number of X-ray counts. For Sample A, when the \feka\ 
emission lines are detected, the total number of counts is at 
least $\sim500$. The \feka\ emission-line luminosity and flux
errors are calculated using the maximum error
estimates. For instance, the upper bound of $L(\mbox{\feka})$ is
calculated using the upper bounds of both EW(\feka) and
$L(\mbox{2--10~keV})$; the upper bound of $f(\mbox{\feka})$ is
calculated using the upper bound of $L(\mbox{\feka})$ and the
lower bound of luminosity distance $d_{\rm L}-\delta d_{\rm L}$.
This uncertainty ignores any systematic uncertainty produced by errors
in the cosmological model. 

For Sample B, because the parameter uncertainties are not given
in GBS08\nocite{gib08}, we simply apply a 20\% uncertainty for all 
luminosity values as a first-order approximation.

%% file: sec4.tex
\section{Drivers of the \civ\ Baldwin Effect}
\label{drive}
\subsection{Comparison with Previous Work}
\label{verify}
We will first examine some important
relations to determine if our measurements
are consistent with previous work.
It has been argued that \aox\ has no detectable redshift dependence
(e.g., Strateva et al. 2005; Steffen et al. 2006; but see Kelly et al. 
2007)\nocite{str05,steffen06,kel07}, 
so in this paper we neglect any redshift evolution of \aox.

Fig.~\ref{fig-civbeffl} displays the plot of EW(\civ) against \luv\
for the combined sample, distinguished by luminosity. We use the monochromatic
luminosity at 2500~\AA\ rather than the traditional BEff 
wavelength (1450~\AA) because our choice is more convenient to 
compare the BEff with the
correlation between \civ\ and \aox. The luminosities at these two wavelengths 
are well correlated. The quantity $f_{1400}/f_{2500}$ is Gaussian distributed 
with a dispersion of $\sigma\sim0.15$ (Fig.~3 in Gibson 
et al. 2009\nocite{gib08b}), so using \luv\ instead of $l_\nu(1450~\mbox{\AA})$
should only add a small dispersion to the data points but will not significantly
affect the slope of the BEff. 
We fit the data points linearly in logarithmic space using the EM 
(Expectation-Maximization) method (Dempster, Laird \& Rubin 1977\nocite{dem77}; 
Table~\ref{tab-fit}). It is clear that EW(\civ) decreases with 
\luv\ (Fig.~\ref{fig-civbeffl}). 

It has been reported that the slope of the BEff becomes steeper for 
high-luminosity quasars. For example, \citet{die02} obtained a 
\civ\ BEff slope ($-0.14\pm0.02$), which was shallower than the 
value reported in previous studies ($-0.22\pm0.05$, Green 1996; also see
Osmer, Porter \& Green 1994; Laor et al. 1995\nocite{osm94,lao95,gre96}). 
Using only the EW(\civ) measurements for their high-luminosity subsample with
$\lambda L_\lambda(1450\mbox{~\AA})\gtrsim10^{44}$ erg~s$^{-1}$,
\citet{die02} obtained a steeper slope of the BEff of $-0.20\pm0.03$ for
\civ. To investigate the slope change with luminosity, we fit our data points
with \luv$<30.5$. We find that the slope of the low-luminosity sample is
consistent with the slope of the entire sample within $1~\sigma$ 
(Table~\ref{tab-fit}). Therefore, although our dataset exhibits a 
suggestive \emph{trend} 
which disagrees with \citet{die02}, we do not find significant
changes of slope over luminosity. 

The large scatter in the BEff could have several causes, including
observational error, intrinsic variation of the BEff (Osmer \& Shields 1999; 
Shields 2007 and references therein)\nocite{osm99,shi07}, 
luminosity dependence of the BEff slope, and the redshift dependence of the 
BEff. Vanden Berk et al.  (2009, in preparation)\nocite{van08} found
that the slope of the BEff does not change across redshift, but its 
scaling factor (or equivalently, the EW of a broad emission line at a fixed
monochromatic luminosity) exhibits a significant change
Our BEff slope is an 
overall average across the redshift and luminosity range we covered; 
remember that there is a strong redshift-luminosity correlation in our sample. 

We examined the \luv-\lx\ and \luv-\aox\ relations for both
Sample~A and the combined sample using survival analysis (ASURV,
Lavalley, Isobe \& Feigelson, 1992\nocite{lav92})
if censored data are involved, and find that all the results are 
consistent with previous work.
When we calculate the slope of the \feka\ BEff, we apply the
Buckley-James method (Table~\ref{tab-fit}, Buckley \& James 1979;
Lavalley, Isobe \& Feigelson 1992) \nocite{buc79,lav92} included
in ASURV to perform linear regression because the EW(\feka) 
contains censored values. The slope of the \feka\ BEff of Sample A 
($-0.115\pm0.062$) is 
consistent with the result from the RQ sample ($-0.102\pm0.052$) in
\citet{jiang06}. 

\subsection{The Effects of \aox\ on the \civ\ and \feka\ BEffs}
Fig.~\ref{fig-ewciv_aox} shows the plot of EW(\civ) against \aox\ for 
the combined sample; the regression result from the EM algorithm for the linear 
relation is 
\begin{equation}
\label{regressew1}
\log{\mbox{EW(C~{\sc iv})}}=(1.035\pm0.075)\alpha_{\rm ox}+(3.301\pm0.119)
\end{equation}
and the Spearman correlation coefficient is 0.607 ($P_0<0.001$).\footnote{$P_0$ is the confidence 
level of the null hypothesis. Therefore, the smaller $P_0$ is
the more likely the correlation exists.} Because \aox\
is an indicator of the hardness of the SED which
controls the ionization level of \civ\ surrounding the central
engine, Fig.~\ref{fig-ewciv_aox} demonstrates that as the
ionizing flux becomes harder (\aox\ increases), the
\civ\ emission has a strong positive response to \aox.

Both \aox\ and \luv\ are correlated with EW(\civ); which is a more
fundamental driver? To investigate this issue, we applied PCA to EW(\civ),
\luv, and \aox\ using the combined sample, Sample A, and a reduced sample.
Table~\ref{tab-pcaresult} presents Pearson, Spearman, and Kendall's
correlation and partial-correlation coefficients (if available), along with
significance levels for these three samples. We present the statistical results
of Sample A for comparing the correlation results of \civ\
and \feka. Because the combined 
sample contains censored data for the \aox\ values, we must use 
survival analysis to calculate the correlation coefficients. However, 
algorithms are not available for calculating \emph{all} the correlation
and partial-correlation coefficients for censored data. For example,
the empty entries in Table~\ref{tab-pcaresult} are due to the unavailability
of corresponding algorithms. In order to compare the changes of 
correlation strength when the third
parameter is controlled, we construct a reduced sample with the censored data
removed, considering that this only excludes a small fraction 
($\sim6\%$) of the entire data set and should not affect the statistical 
properties of the sample. We can see that the \civ\ BEff is significantly
weakened when \aox\ is held fixed; the correlation coefficient
drops from $-0.580$ to $-0.224$ (Spearman). On the other hand, the 
correlation coefficient between
EW(\civ) and \aox\ also drops significantly when \luv\ is held
fixed, from 0.615 to 0.332. This implies that both \aox\ and \luv\ are 
driving the change of EW(\civ).

The \feka\ BEff plot (not shown) of our sample (Sample A) is very similar to the
correlation shown of Fig.~4 in \citet{jiang06} except that our sample 
size is smaller. Fig.~\ref{fig-ewfeka_aox} shows EW(\feka) plotted against
\aox. The Spearman test gives a much weaker correlation coefficient 
($-0.230$ with $P_0=0.111$) than that for \civ\ ($-0.304$ with 
$P_0<0.001$). In addition, simple $\chi^2$ fitting 
produces a slope of $0.046\pm0.154$, consistent with zero. 
The consistency between the correlation analysis and regression
result provides strong evidence that EW(\feka) is not correlated
with \aox.

\subsection{Effects of AGN Variability on BEff Relation Scatter}
Because of the ubiquity of AGN variability, combined with the
different times of the optical and X-ray observations, our values
of \aox\ do not reflect the spectral hardness at a specific time
but are randomly distributed around their mean values. 
The deviation of \aox\ from its mean value would be
$\sim0.083$, assuming that the variation amplitudes are 30\% for
\luv\ and 40\% for \lx\ (e.g., Strateva et al. 2005, 
GBS08\nocite{str05,gib08}). 

To check how much of the scatter of our correlations could be attributed to
variability, we performed two simple tests on our combined sample. We follow 
the method used in \S~3.1 of GBS08\nocite{gib08} and introduce 
\dew, which is the difference between the observed EW(\civ) 
and the EW calculated from linear regression (Eq.~\ref{regressew2}), i.e.,
$\Delta\log{\mbox{EW}}=\log{\mbox{EW}}-\log{\mbox{EW}(l_\nu(2500~\mbox{~\AA}))}$. We define $\mu$ and $\sigma$ as the mean and dispersion of the distribution
of \dew. To calculate $\mu$ and $\sigma$, we maximize the likelihood function
\citep{mac88}:
\begin{equation}
  \label{lf}
  L = \prod_{i}\frac{1}{\sqrt{2\pi\left(\sigma_i^2+\sigma^2\right)}}\exp{\left[-\left(\Delta\log{\mbox{EW}_i}-\mu\right)^2/2(\sigma_i^2+\sigma^2)\right]}
\end{equation}
in which the subscript $i$ represents each object and $\sigma_i$ is the 
uncertainty of \dew\ associated with each $\Delta\log{\mbox{EW}}_i$. The 
maximization of $L$ requires $\mu=0.01$ and $\sigma=0.23$. 

Next, we estimate the potential scatter of $\Delta\log{\mbox{EW}}$ due to
variability. We need to consider two terms: $\log{\mbox{EW}}$ and 
$\log{\mbox{EW}}(l_\nu(2500~\mbox{\AA}))$. To first approximation, 
$\mbox{EW}\sim f_{\rm line}/f_{\rm cont}$ in which $f_{\rm line}$ is the 
emission-line flux and the $f_{\rm cont}$ is the continuum flux. 
The emission-line variability of six luminous 
quasars at $z=2.2$--$3.2$ was recently reported by \citet{kas07}. 
The mean fractional 
variation $F_{\rm var}$\footnote{$F_{\rm var}$ is defined as the RMS of the 
intrinsic variability relative to the mean flux \citep{rod97}.} is 
$\approx0.096$ by averaging the fractional variation of \civfull\ of all 
six quasars. The \civ\ emission-line variability of a number of Seyfert galaxies
has been studied, including \object{Fairall~9} \citep{rod97}, \object{NGC~5548}
\citep{cla91}, \object{NGC~7469} \citep{wan97}, \object{NGC~3783} \citep{rei94},
and \object{3C~390.3} \citep{obr98}. By averaging the fractional variations of
Seyferts and quasars above, we obtain an \emph{average} emission-line variation
$\langle F_{\rm var}\rangle=0.130$. The typical variation of 
$l_\nu(2500~\mbox{\AA})$ is $\sim30\%$ (e.g., Strateva et al. 
2005)\nocite{str05}. Therefore, the scattering of $\log{\mbox{EW}}$ 
contributed from variability is \emph{estimated} (assuming all independent
variables are Gaussian distributed) as 
$$\sigma\left(\Delta\log{\mbox{EW}}\right)=\frac{1}{\ln{10}}\sqrt{\left(\frac{\delta f_{\rm line}}{f_{\rm line}}\right)^2+\left(\frac{\delta f_{\rm cont}}{f_{\rm cont}}\right)^2+a^2\left[\frac{\delta l_\nu(2500~\mbox{\AA})}{l_\nu{2500~\mbox{~\AA}}}\right]^2}\approx0.144.$$
In the calculation above, $a=0.198$ (Eq.~(\ref{regressew2})). This exercise 
indicates that at least 60\% of the scatter around the BEff in our sample 
can be attributed to AGN variability.

We performed a similar test for the \aox-EW(\civ) relation. Because the set of 
\aox\ contains censored data, we can only use the reduced data set (258 
objects).
The maximization of 
$L$ (Eq.~\ref{lf}) yields $\mu=0.01$ and $\sigma=0.22$. The potential
dispersion of this relation assuming all scatter comes from variability is
\emph{estimated} (assuming Gaussian distributions) as
$$\sigma\left(\Delta\log{\mbox{EW}}\right)=\frac{1}{\ln{10}}\sqrt{\left(\frac{\delta f_{\rm line}}{f_{\rm line}}\right)^2+\left(\frac{\delta f_{\rm cont}}{f_{\rm cont}}\right)^2+(0.3838a)^2\left[\left(\frac{\delta l_\nu(2500~\mbox{\AA})}{l_\nu(2500~\mbox{~\AA})}\right)^2+\left(\frac{\delta l_\nu(2~\mbox{keV})}{l_\nu(2~\mbox{~keV})}\right)^2\right]}\approx0.163.$$
In the calculation above, $a=1.035$ (Eq.~\ref{regressew1}).
This indicates that variability produces at least 75\% of the scatter around
the \aox-EW(\civ) relationship.

In summary, the above two tests demonstrate that a substantial fraction, if not the majority, of the scatter in the correlations above can be attributed to 
X-ray and UV/optical variability. It should be possible to make the 
correlations tighter if the UV/optical and X-ray data are observed 
simultaneously. 

\subsection{Regressions of EW and Luminosity} The BEff 
provides a potential avenue to infer the luminosity of a
quasar from emission-line observations. Type Ia supernovae
(SNe) are treated as classical standard candles
(e.g., Phillips 1993; Burrows 2000)\nocite{phi93,bur00}, but 
only a few are observed beyond $z\sim1.5$.
If quasars, which are much easier to detect than SNe and can be
observed to much a higher redshift, can be used as standard 
candles, they would
be an important tool for cosmological studies. Soon after the
discovery of the BEff, many investigations considered the
possibility of treating emission-line EW as a luminosity indicator
(e.g., Baldwin 1977b; Wampler 1980)\nocite{bal77b,wam80}.
Unfortunately, the \civ\ BEff usually has a large scatter
\citep{osm99,shi07}; given the small slope in the
$\log{\mbox{EW}}$-$\log{l_\nu}$ plot (on the order of $-0.2$),
the predicted luminosities are very inaccurate. It has
also been shown that the \civ\ BEff is redshift dependent
(Francis \& Koratkar 1995; Vanden Berk et al. 2009, in 
preparation)\nocite{fra95,van08}, making it a less valuable 
probe of cosmology.

Because we are focusing on the influence of \aox\ at the
moment, and will put the redshift factor aside, we will concentrate on
the issue of whether the scatter can be reduced if
we regress EW(\civ) with \luv\ and/or \aox, and if
the prediction of luminosity can be made more accurate
with this approach. The linear-regression
results of EW(\civ) with \aox\ are already shown in Eq.(\ref{regressew1}). 
Similar regressions from \luv\ and both of \luv\ and \aox, 
using the fully parametric EM algorithm, are
\begin{align}
\label{regressew2} 
\log{\mbox{EW(C~{\sc iv})}}&=
	(-0.198\pm0.015)\log{l_{\nu}(2500\mbox{~\AA})}+
	(7.764\pm0.461)\\ 
\label{regressew} 
    \log{\mbox{EW(C~{\sc iv})}}&=
	(-0.107\pm0.021)	\log{l_\nu(2500\mbox{~\AA})}+
	(0.615\pm0.106)		\alpha_{\rm ox}+
	(5.944\pm0.536). 
\end{align}
in which EW(C~{\sc iv}) is in \AA\ and \luv\ is in
erg~s$^{-1}$~Hz$^{-1}$. The last regression was performed on
the combined sample without the censored data (258 objects) because 
the EM algorithm in ASURV does not allow both independent variables to contain
censored data points. 

To evaluate the scatter, we subtract the predicted EW values calculated using
the above equations from the observed values and compute the RMS values of  
the residuals. We see a slight improvement of the RMS values, 
using \aox\, and both \luv$+$\aox\ (Table~\ref{tab-rmsew}). 
The last regression result (Eq.(\ref{regressew})) is consistent with 
Eq.(11) in GBS08\nocite{gib08}, indicating that at least
part of the scatter of the BEff is due to \aox. 

Next, we regress luminosity against EW(\civ) and/or \aox, using the combined 
sample. The results are
\begin{align}
\label{regressl1}
\log{l_\nu(\mbox{2500\ \AA})}&=
	(-1.980\pm0.154)	\log{\mbox{EW(C~\sc iv)}}+
	(34.076\pm0.256)\\ 
\label{regressl} 
\log{l_\nu(\mbox{2500\ \AA})}&=
	(-0.852\pm0.168)	\log{\mbox{EW(C~{\sc iv})}}
	-(2.864\pm0.262)	\alpha_{\rm ox}+
	(27.668\pm0.635). 
\end{align}
The last regression was performed on the censored-data-excluded
sample. We then calculate the RMS values of the residuals after
subtracting the predictions from the equations above
(Table~\ref{tab-rmsew}). The RMS value shrinks by 18\% using 
EW(\civ)$+$\aox, compared to using EW(\civ) alone. We use the standard $F$-test
to check if the two sets of residuals have consistent variance. The 
testing gives
an $F$-statistic of 1.48 with a significance $0.002$, indicating that these
two sets of residuals have significantly different variances and 18\% is
a statistically significant improvement. To use quasars as standard
candles via the BEff, we should at least confine the luminosity within an
uncertainty of 30\%, or, equivalently, $\mbox{rms}<0.1$. 
This cannot be achieved using our current dataset and controlled parameters.

%% file: sec5.tex
\section{Relation Between \feka\ and \civ}
\label{relation} \feka\ is important in AGN studies because it is
the strongest emission line appearing in the X-ray band.
However, the strength of this emission line varies significantly
from object to object, and the line is not detected in most X-ray observations
of quasars. Given this lack of direct observational measurement, it would be
useful to develop a way to predict the expected strength of the line 
empirically.

The EW(\civ) and EW(\feka) do not exhibit a significant correlation 
(Fig.~\ref{fig-ew_feciv}) in Sample A; the correlation has a low 
Spearman rank-correlation 
coefficient ($0.319$ with $P_0=0.027$).
Although the EWs of the two lines are not well correlated, their 
luminosities and fluxes are strongly correlated 
(Fig.~\ref{fig-feciv}), 
with Spearman correlation coefficients $0.529$ ($P_0<0.001$) and $0.551$ 
($P_0<0.001$), respectively (Table~\ref{tab-cfeciv}). The linear 
correlations regressed in
Fig.~\ref{fig-feciv} are
\begin{align}
  \label{l_feciv}
  \log{L(\mbox{Fe~K$\alpha$})}&=(0.588\pm0.079)\log{L(\mbox{C~{\sc iv}})}+(16.164\pm3.416)\\
  \log{f(\mbox{Fe~K$\alpha$})}&=(0.978\pm0.188)\log{f(\mbox{C~{\sc iv}})}-(2.082\pm2.228).
\end{align}

One must always question the significance of relations such as (\ref{l_feciv})
because even if there is no correlation between the observed fluxes of the 
lines, the fact that the line luminosities  for a given object contain the 
same distance factor will 
introduce an apparent correlation in the luminosities. To investigate whether 
this effect is important for our study, we perform a test in which we conduct
correlation and regression analysis for a sub-sample of Sample A. Objects in 
this sub-sample have similar redshifts, and thus they have approximately the
same distances. First, we use objects with $0.06<z<0.09$ because 
this redshift bin
contains a large number of objects. This sub-sample contains 10 objects. 
The correlation coefficient of 
$f(\mbox{Fe~K}\alpha)$-$f(\mbox{C~{\sc iv}})$ is 0.624 ($P_0=0.061$) and of 
$L(\mbox{Fe~K}\alpha)$-$L(\mbox{C~{\sc iv}})$ is 0.709 ($P_0=0.033$). The
regression results are
\begin{align}
  \label{l_feciv_sub}
  \log{L(\mbox{Fe~K$\alpha$})}&=(0.748\pm0.200)\log{L(\mbox{C~{\sc iv}})}+(9.330\pm8.702)\\
  \log{f(\mbox{Fe~K$\alpha$})}&=(0.771\pm0.241)\log{f(\mbox{C~{\sc iv}})}-(4.280\pm2.819).
\end{align}
Both the correlation and regression results of this sub-sample are consistent 
with the results of the entire sample within uncertainties, suggesting that
the luminosity correlation of the \civ\ and \feka\ lines may be real and is  
not a consequence of multiplying the two fluxes of a given object by the same 
large distance factor.  The luminosity of \civ\ emission line increases 
faster than the luminosity of \feka.

One might be concerned that the correlation between $L$(\feka) and $L$(\civ) 
is \emph{artificial} because the calculation of the \feka\ measurements
involves $L$(2--10~keV) (Eq.~(\ref{flfeka})), 
which is proportional to 
\lx\ (Eq.~(\ref{l2kevlx})), and \lx\ is correlated with \luv, which is
correlated with EW(\civ), i.e., the \civ\ BEff. EW(\civ) is calculated from 
continuum and emission line luminosity, so apparently, the $L$(\feka) and 
$L$(\civ) are not independent before we perform the correlation. However, 
our calculation of \feka\ is simply reversing the $F$(2--10~keV)/EW calculation
of \citet{jiang06}, so $F$(\feka) is not actually dependent of $F$(2--10~keV)
and hence \luv. In essence, we have the values of $F$(\feka) independent of 
$F$(\civ). Therefore, our $L$(\feka) and $L$(\civ) correlation, which is 
\emph{expected} from existing relations, is not an artifact correlation.

That the EWs of \civ\ and \feka\ are uncorrelated is consistent
with the result by \citet{pag04a} and further demonstrates that the
\civ\ and \feka\ emission lines are unlikely to have the same
origin. This result is not surprising because \feka\ and \civ\ are produced
in different processes. The correlation between their luminosities 
is probably a combination of effects between their EWs (uncorrelated) and
continuum (strongly correlated). The flux correlation, although
empirical and not very tight, is a useful first order
estimation of the \feka\ line flux given the UV spectra in the
rest-frame of an AGN.

%% file: sec6.tex
\section{Discussions and Conclusions}
\label{conclusions} 
We have compiled a sample of 272 Type
1 AGNs and quasars that have UV and X-ray measurements, among which 
\feka\ emission lines are detected in 50 objects. The sample covers a wide
range of redshift ($0.009\lesssim z\lesssim4.720$), and a wide range of 
luminosity from Seyfert galaxies to the most-luminous quasars 
($27.81\lesssim\log{l_\nu(2500~\mbox{\AA})}\lesssim33.04$). These properties
allow us to study the overall properties of AGNs rather than focusing on a
particular redshift or luminosity. It also has a high X-ray detection rate 
($\sim96\%$), which lets us obtain robust statistics. We have performed 
correlation and regression analyses using this sample and draw the 
following conclusions:
\begin{enumerate}
  \item The \civ\ BEff is driven by both \aox\ and \luv, or equivalently,
  by \lx\ and \luv. This implies that changes in the ionizing flux induce
  changes in the ionization state of the BELR, producing more \civ\
  ions when the SED becomes harder and vice versa. This is supported both by 
  correlation and regression anlayses:
  \begin{itemize}
    \item The partial correlation 
  between EW(\civ) and \luv\ when \aox\ is controlled is weaker  
  than the regular correlation between EW(\civ) and \luv.
    \item The scatter in the linear regression decreases when we 
  regress EW with \aox+\luv\ compared with \luv\ alone.
  \end{itemize}
  Although the
  reduction of the scatter due to adding another regression parameter is not
  sufficiently large to treat quasars as standard candles, it demonstrates that 
  a significant fraction of the scatter attributes to \aox, and can be reduced
  by including it in regression analysis.
  \item EW(\feka) exhibits no strong correlation
  with either \aox\ or EW(\civ). This implies that \feka\ is not likely to have
  the same origin as \civ. 

  \item There may be a correlation between the luminosities
  of \feka\ and \civ\ with a logarithmic slope of $0.588\pm0.079$.
  This correlation is possibly because both of these two quantities 
  involve
  a factor related to the scale of the line emitting
  regions and the slope indicates that the \civ\ emission line luminosity
  increases faster than the \feka. 
\end{enumerate}
Although \aox\ is a fundamental influence on EW(\civ), there is still
a significant scatter in the EW(\civ)-\aox\ diagram. As we have
demonstrated, most of the scatter is contributed by variability, but 
another likely contribution source is the nature of \aox\ which only connects 
the flux points at 2500~\AA\ and 2~keV but misses the Big Blue Bump,
which is expected to play an important role in the photoionization
process. The shape of an AGN SED can be very different depending on
Eddington ratio ($L_{\rm bol}/L_{\rm Edd}$) but still have 
a fairly constant \aox\ \citep{vas07}. It is perhaps more appropriate 
to use a point near $\sim250$~\AA\ 
instead of 2500~\AA\ to calculate a revised \aox\ \citep{she08}. This new 
\aox\ might be more strongly correlated with EW(\civ). However, this 
requires challenging observations that cannot be achieved at present 
for most AGNs.  

%% file: figure.tex
\begin{figure}
\centering
\includegraphics[angle=90,width=16cm]{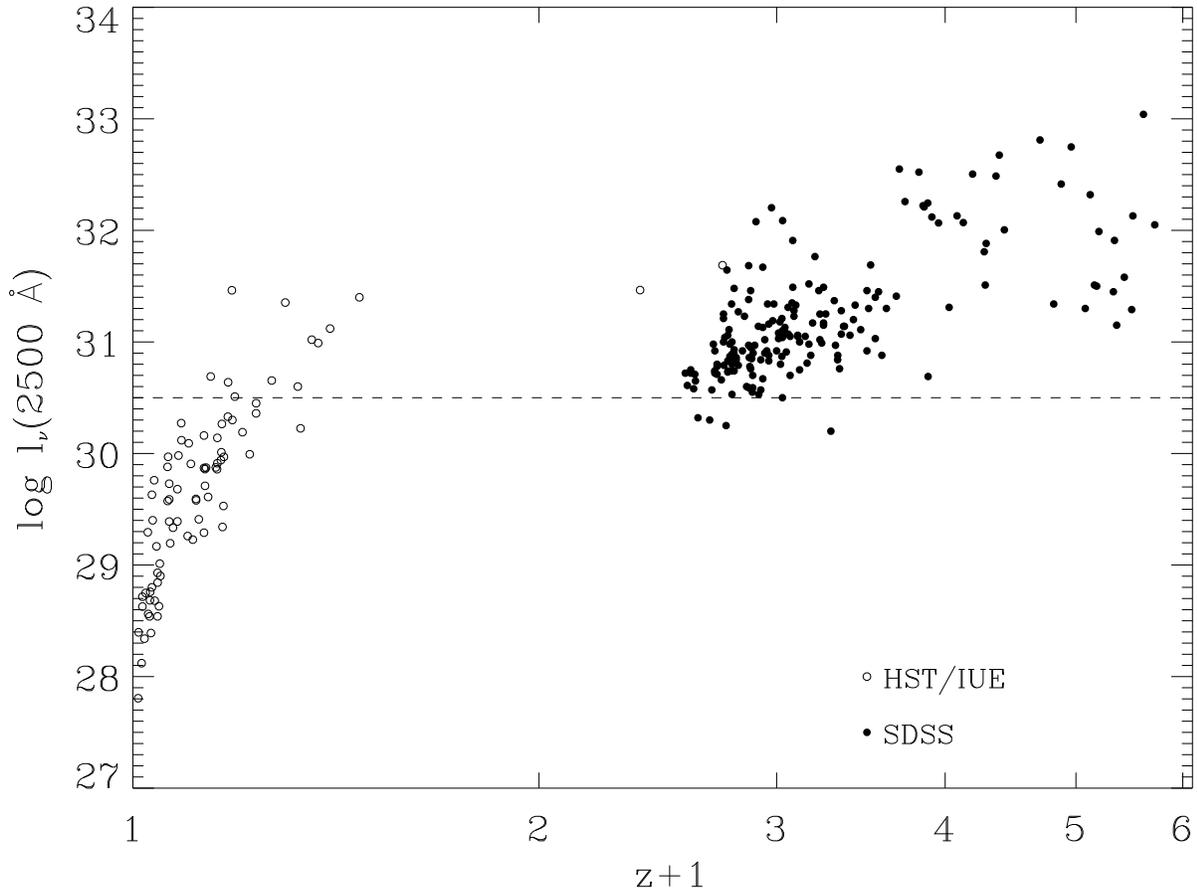}
\caption{The luminosity-redshift diagram of the combined sample containing 272
objects distinguished by observational facilities. The gap for 
$0.5\lesssim z\lesssim1.5$ arises because of the detection limit of \emph{HST}/\emph{IUE} 
and wavelength coverage of the SDSS. The dashed line marks the position 
where \luv$=30.5$. 
\label{fig-lz}}
\end{figure}

\begin{figure}
  \begin{center}
    \begin{tabular}{c}
      \includegraphics[width=6cm,angle=90]{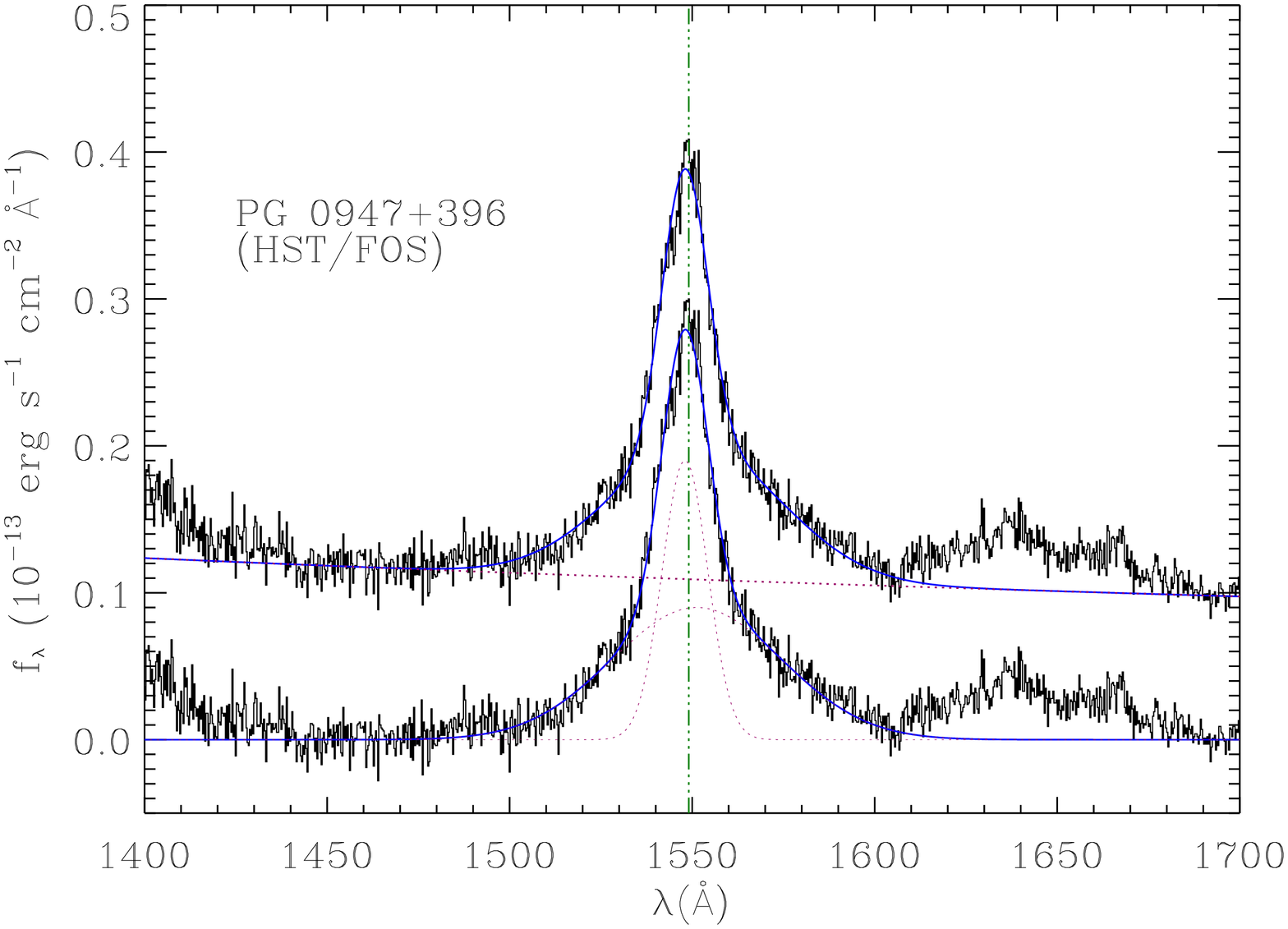}\\
      \vspace{2mm} \includegraphics[width=6cm,angle=90]{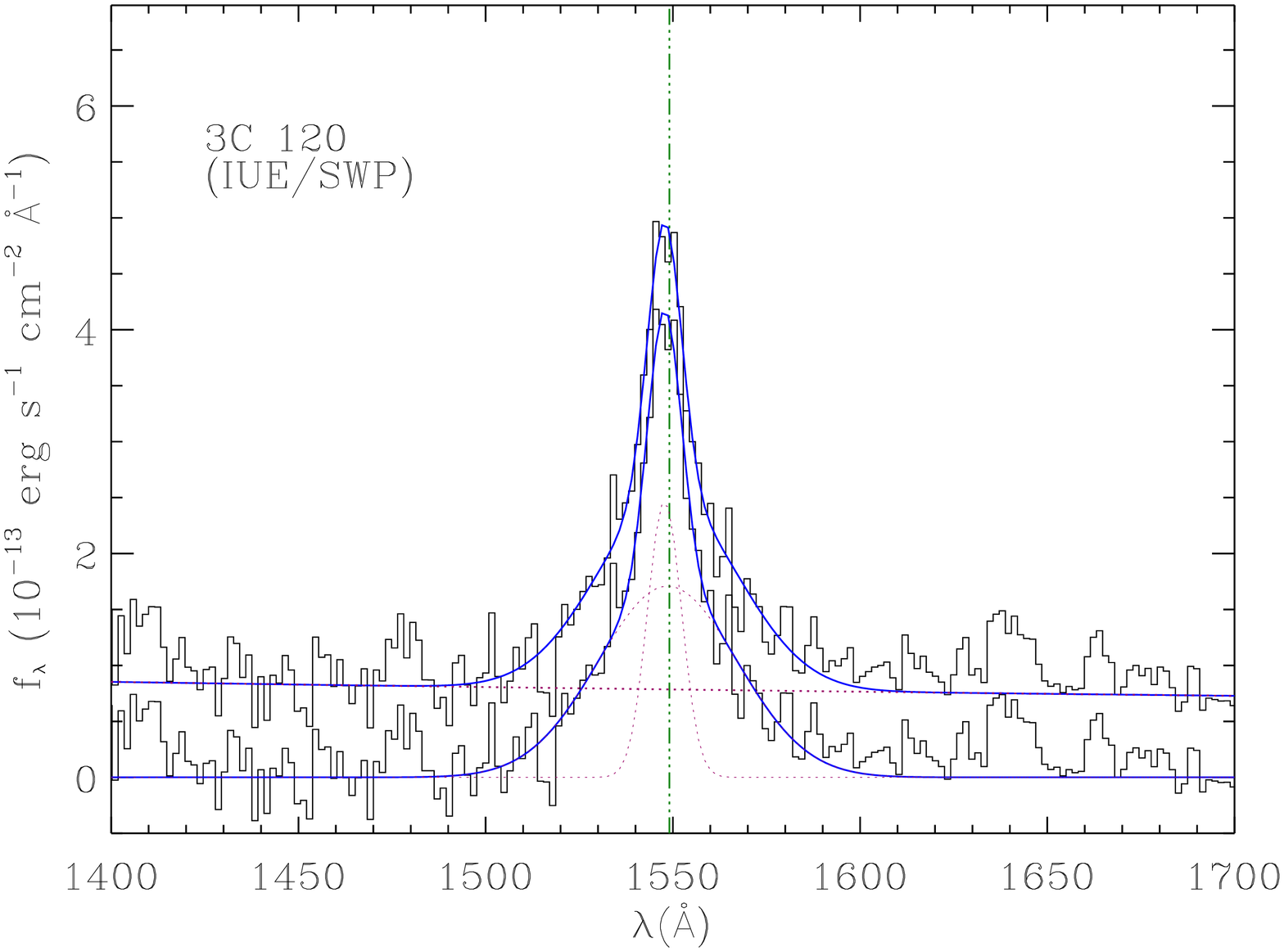}\\
      \vspace{2mm} \includegraphics[width=6cm,angle=90]{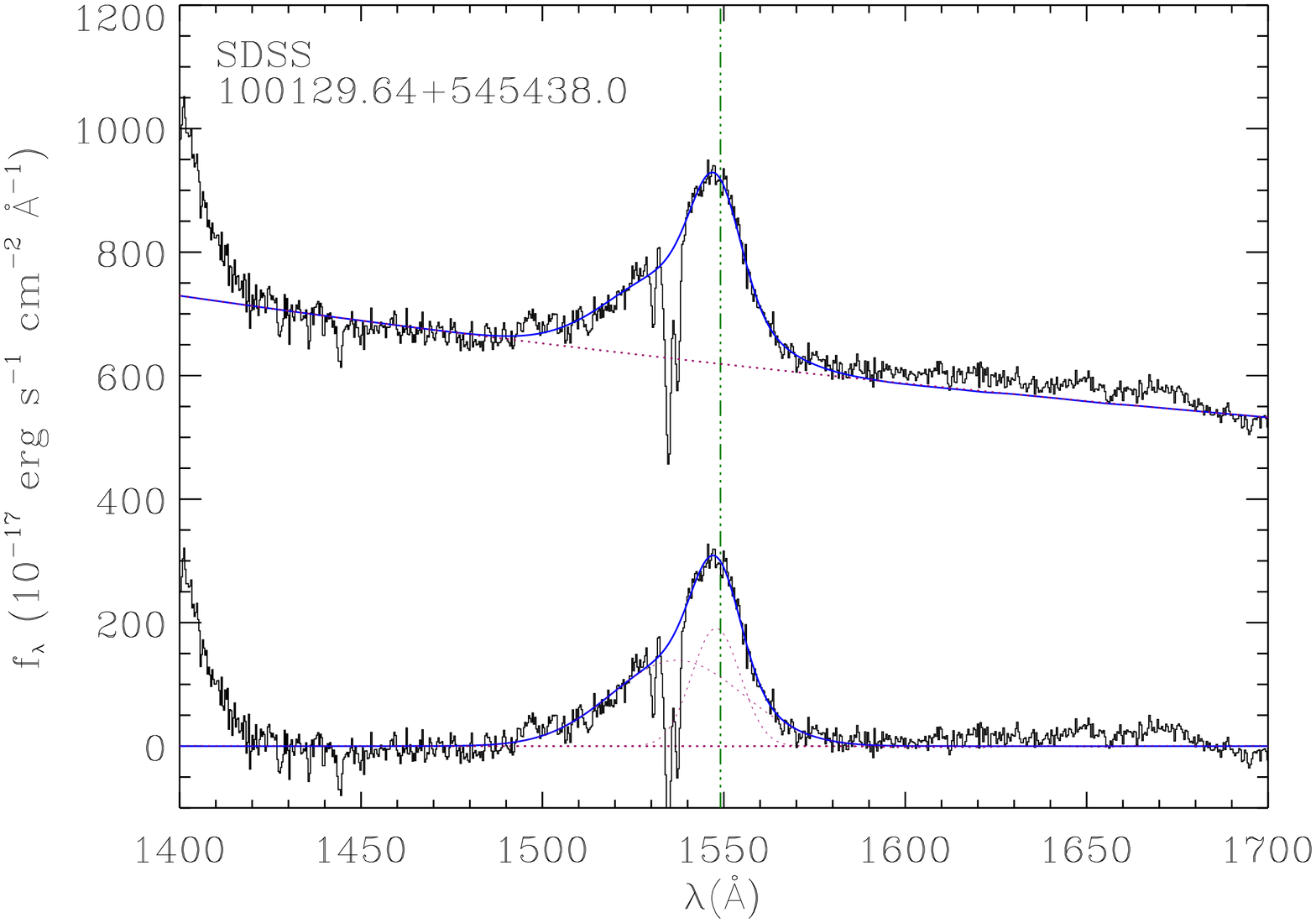}
    \end{tabular}
    \caption{Continuum and \civ\ emission-line fit examples. In each panel,
             the upper spectrum is the original and the lower spectrum is 
             continuum subtracted; blue solid curves 
             are the fits to the spectra.
             The cyan dotted curves are emission-line components. For 
             PG~0947$+$396 and 3C~120, these components include the power-law 
             continua and two Gaussian profiles for \civ. For 
             SDSS~J100129.64$+$545438.0, we also plot the
             iron emission forest and small Balmer bump components, although
             they are so weak that they are almost invisible. The rest-frame 
             spectral resolutions of these spectra are, from top to bottom, 
             $\sim1.6$~\AA, $\sim5$~\AA, $\sim0.8$~\AA. \label{fig-fit}}
  \end{center}
\end{figure}
\clearpage

\begin{figure}
\begin{center}
  \includegraphics[angle=90,width=16cm]{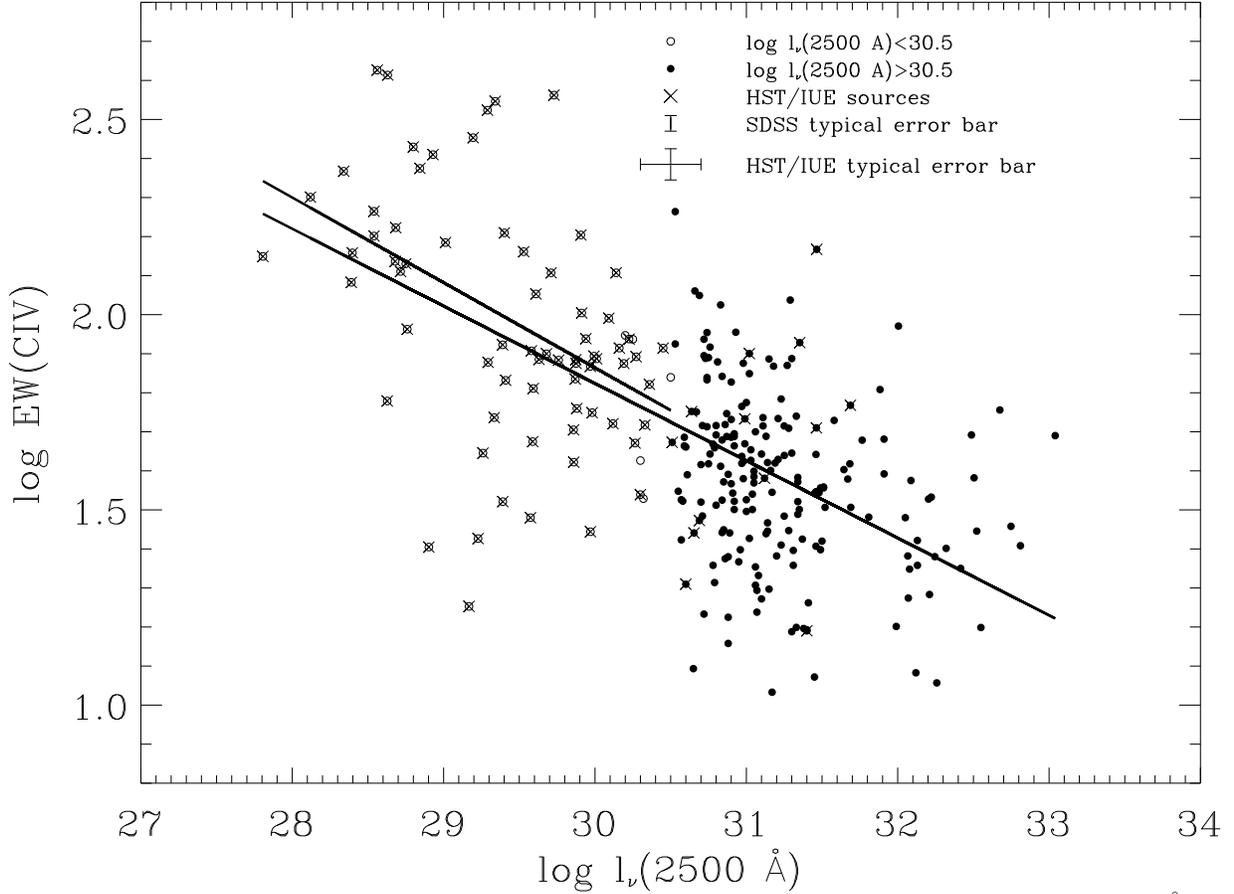}
\caption{The \civ\ BEff diagram for the combined sample. The EW(\civ) is 
given in \AA, and \luv\ is in units of \luvunit. Points are 
distinguished by luminosity; the short solid line is the best linear fit to 
low-luminosity points and the long one is the best linear fit 
to the entire sample.  
\label{fig-civbeffl}}
\end{center}
\end{figure}
\clearpage

\begin{figure}
\begin{center}
\includegraphics[angle=90,width=16cm]{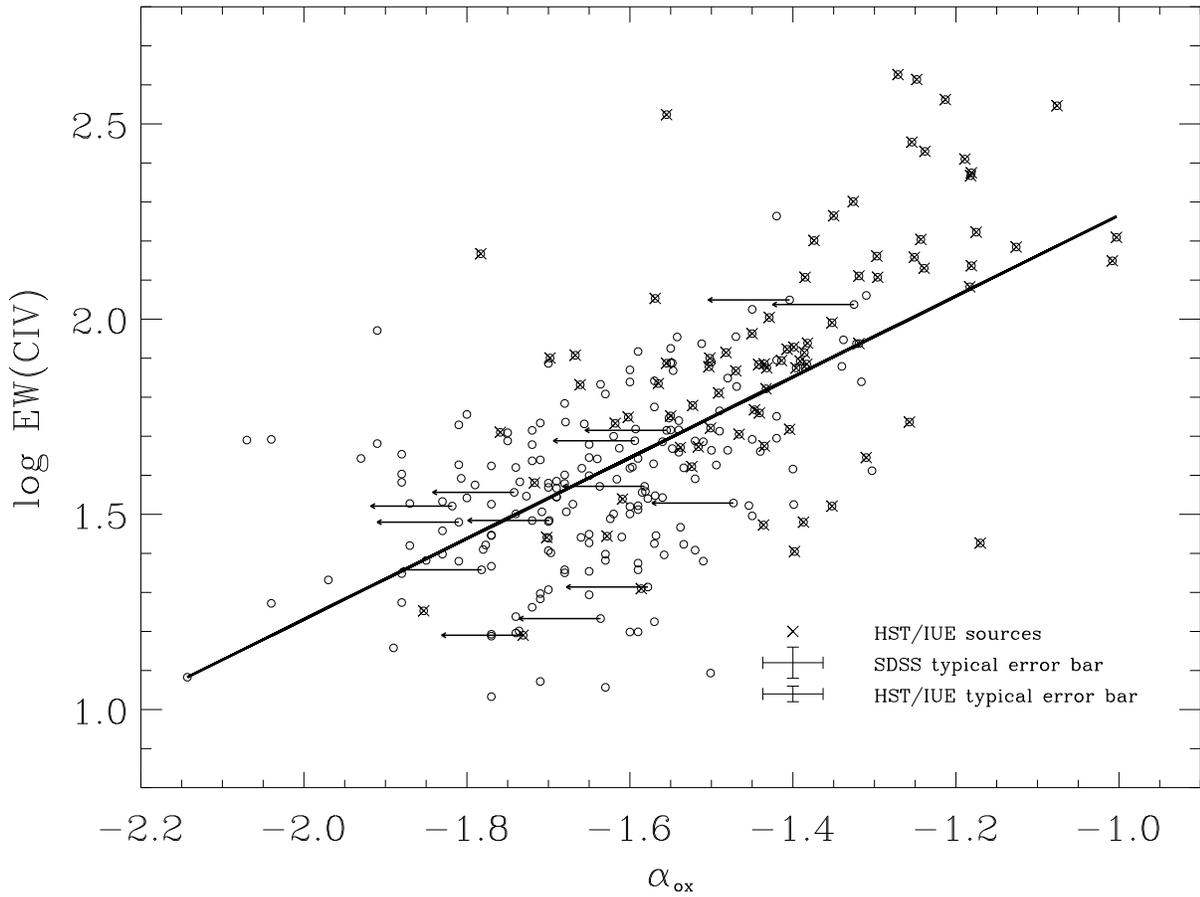}
\caption{The correlation between EW(\civ) and \aox.  Upper limits on 
\aox\ are marked with arrows. The solid line is the best linear fit using the 
EM algorithm. 
\label{fig-ewciv_aox}}
\end{center}
\end{figure}
\clearpage

\begin{figure}
\begin{center}
\includegraphics[angle=90,width=16cm]{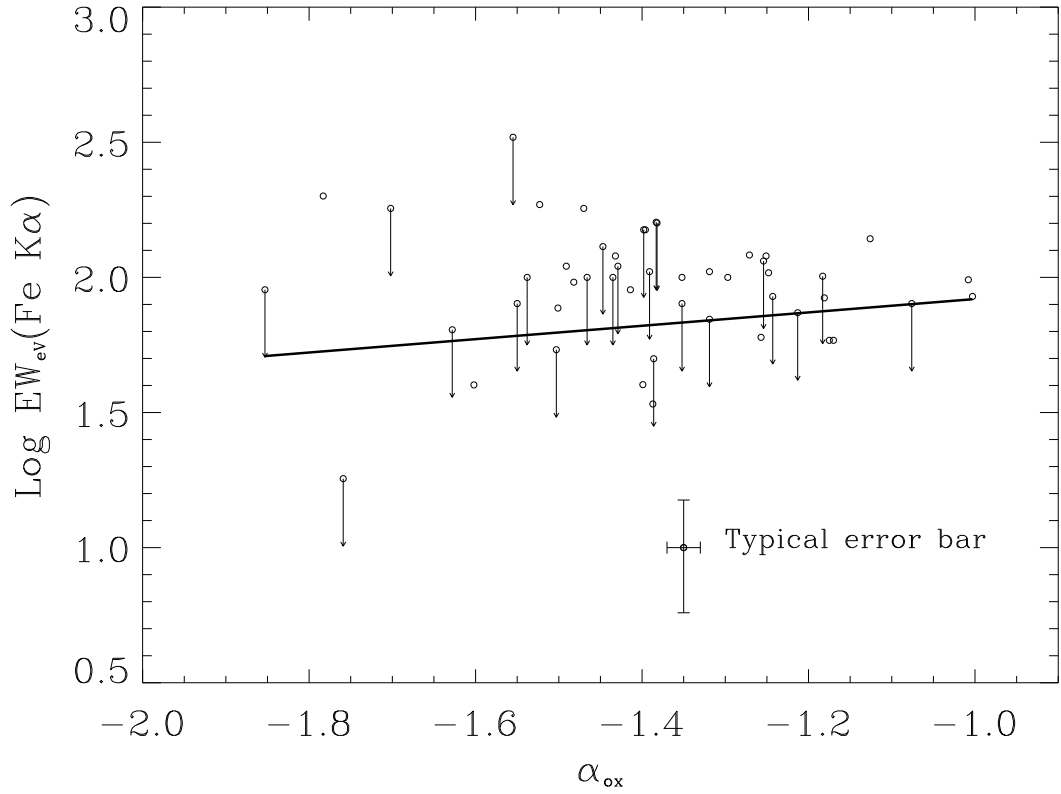}
\caption{The correlation between EW(\feka) and \aox\ for the core
sample. The solid line is a linear fit using the ASURV package for
the censored data. The typical error bar of the data is displayed
at the top-right corner of the plot. \label{fig-ewfeka_aox}}
\end{center}
\end{figure}
\clearpage

\begin{figure}
  \begin{center}
  \includegraphics[angle=90,width=16cm]{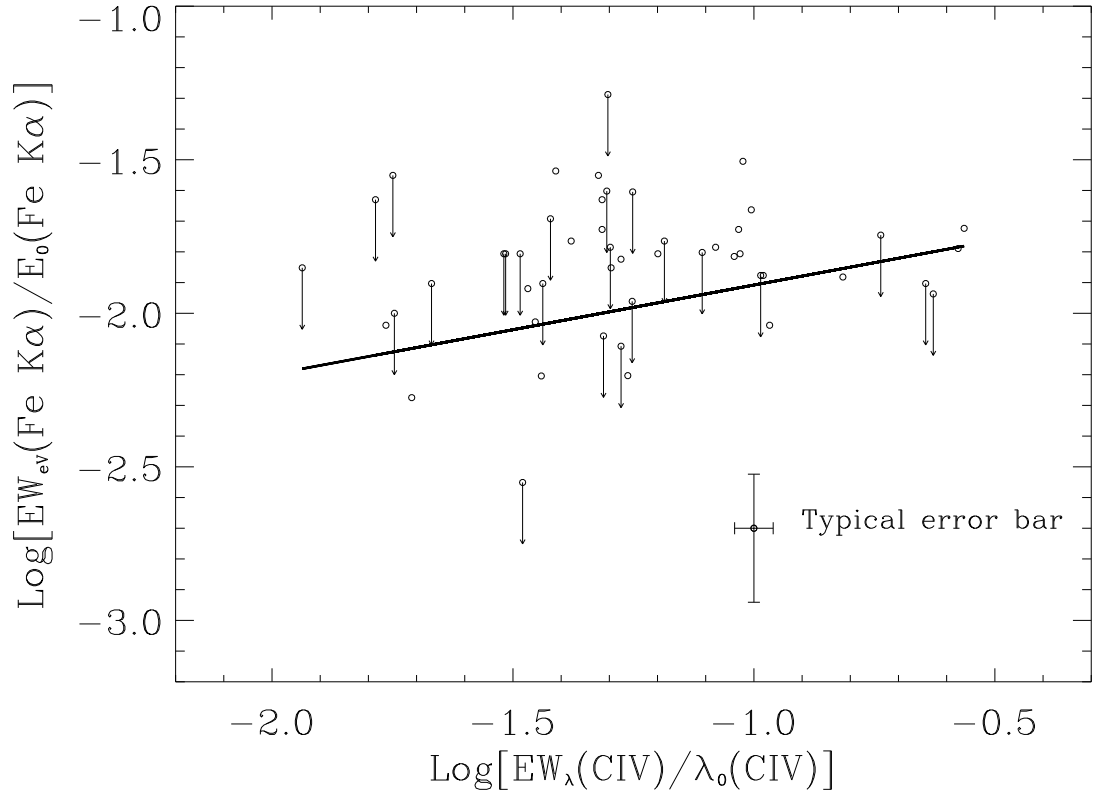}
  \caption{Plot of
          $\log{\left[\mbox{EW(\feka)}/\mbox{E}_0(\mbox{\feka})\right]}$ vs.
          $\log{\left[\mbox{EW(\civ)}/\lambda_0(\mbox{\civ})\right]}$ for the
          core sample. Because the units of EW(\feka) and EW(\civ) are 
          different, we divide them by central energy 
          $\mbox{E}_0(\mbox{Fe~K}\alpha)=6.4\mbox{~keV}$ and central wavelength 
          $\lambda_0(\mbox{C~{\sc iv}})$ to make them dimensionless.
  \label{fig-ew_feciv}}
  \end{center}
  \end{figure}
\clearpage

\begin{figure}
  \begin{center}
    \begin{tabular}{c}
       \includegraphics[angle=90,width=12cm]{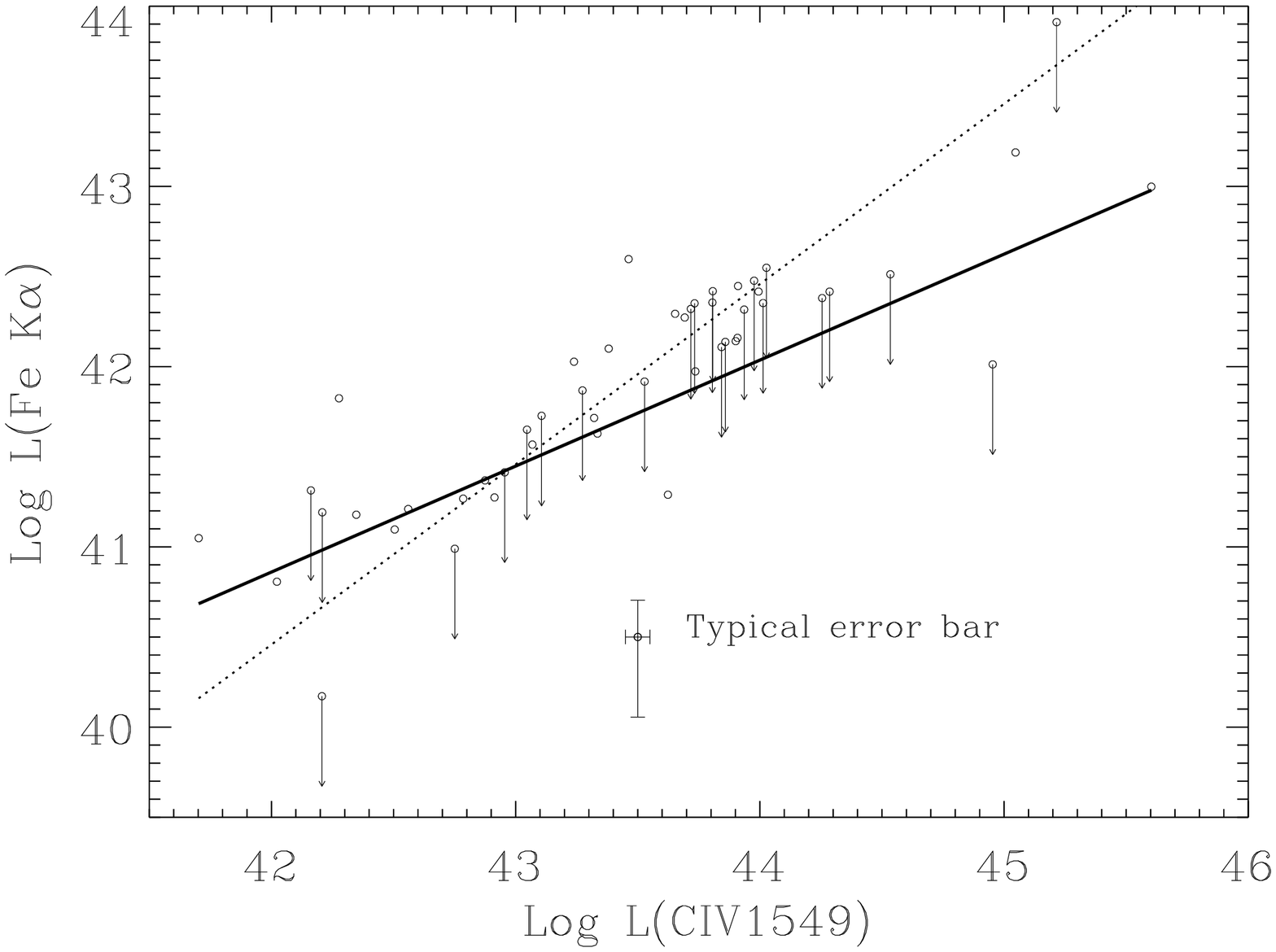}\\
       \includegraphics[angle=90,width=12cm]{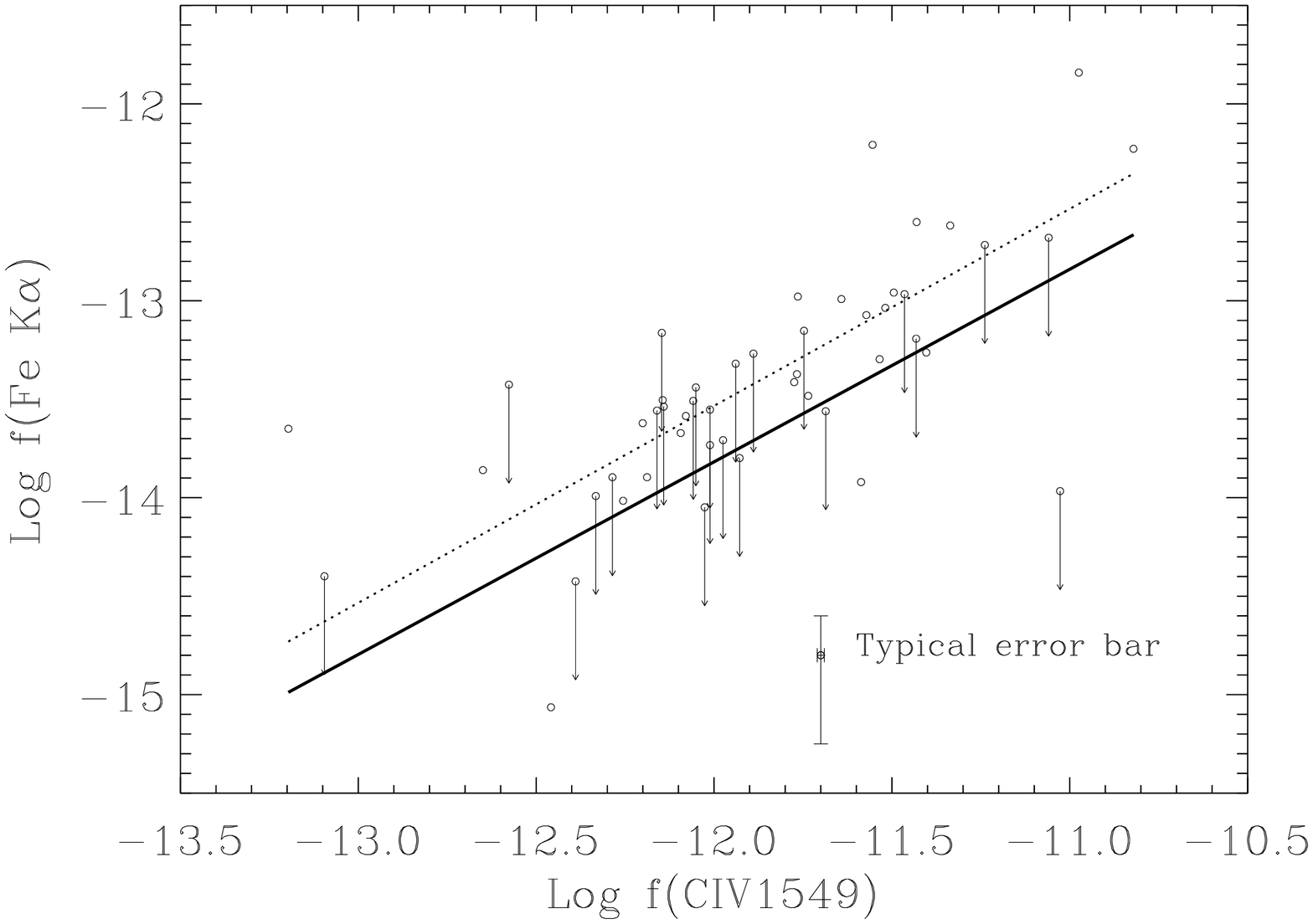}
    \end{tabular}
  \caption{Plot of $\log{L(\mbox{Fe~K}\alpha)}$ vs. $\log{L(\mbox{\civ})}$
          (upper panel) and $\log{f(\mbox{\feka})}$ vs. $\log{f(\mbox{\civ})}$ 
          (lower panel) for the core sample. Upper limits are denoted as 
          downward arrows. The solid lines are the best linear fits to the data 
          using ASURV. For comparison, we show the best linear fits with a unity
          slope in dotted lines. The data symbols and typical error bars are 
          labelled at the bottom right corners.
  \label{fig-feciv}}
  \end{center}
\end{figure}
\clearpage

%% file: table.tex
\begin{deluxetable}{lrrrrrr}
\tablecolumns{5} \tabletypesize{\tiny}
\tablecaption{Summary of samples.\label{tab-samsum}}
\tablewidth{0pc}
\tablehead{
\colhead{Sample} &
\colhead{SDSS}   & 
\colhead{HST}    &
\colhead{IUE}    & 
\colhead{Total}  &
\colhead{Redshift range} &
\colhead{$\log{l_\nu(2500~\mbox{\AA})}$ range}}
\startdata
A & 0 & 34 & 16 & 50 & 0.009--1.735 & 27.81--31.69\\
B & 98 & 0 & 0 & 98 & 1.7--2.7 & 30.53--31.67\\
C & 91 & 13 & 20 & 124 & 0.015--4.720 & 28.12--33.04\\
\hline
Combined & 189 & 47 & 36 & 272 & 0.009--4.720 & 27.81--33.04
\enddata
\end{deluxetable}

\enlargethispage*{1in}
\begin{deluxetable}{lccccccc}
\tablecolumns{8} \tabletypesize{\tiny} \tablecaption{UV properties
of the combined sample.
\label{tab-sampleu}} \tablewidth{0pc} \tablehead{
  \colhead{Object}                          &
  \colhead{$z$}                             &
  \colhead{$\log{l_\nu(2500\mbox{\ \AA})}$} &
  \colhead{EW(\civ)}                        &
  \colhead{\aox}   	                    &
  \colhead{Flag\tablenotemark{1}}           &
  \colhead{UV/Optical}                      &
  \colhead{Sample}                         \\

  \colhead{}                                &
  \colhead{}                                &
  \colhead{(\luvunit)}                      &
  \colhead{(\AA)}                           &
  \colhead{}                                &
  \colhead{}                                &
  \colhead{Instrument}                      &
  \colhead{}
}
\startdata
NGC4593 & 0.009 & $27.81\pm 0.15$ & $141.0\pm 23.2$ & $-1.008\pm 0.066$ & 0 & HST & A\\
NGC3783 & 0.010 & $28.40\pm 0.08$ & $144.0\pm  9.4$ & $-1.251\pm 0.045$ & 0 & HST & A\\
Mkn352 & 0.015 & $28.12\pm 0.15$ & $200.0\pm 12.9$ & $-1.326\pm 0.067$ & 0 & IUE & C\\
Mrk1044 & 0.016 & $28.63\pm 0.04$ & $ 60.1\pm  1.3$ & $-1.523\pm 0.037$ & 0 & HST & A\\
NGC7469 & 0.016 & $28.72\pm 0.28$ & $129.0\pm 25.1$ & $-1.319\pm 0.111$ & 0 & HST & A\\
MCG8$-$11-11 & 0.020 & $28.34\pm 0.19$ & $233.0\pm 39.1$ & $-1.182\pm 0.082$ & 0 & IUE & C\\
Mkn79 & 0.022 & $28.75\pm 0.09$ & $135.0\pm 11.3$ & $-1.239\pm 0.049$ & 0 & IUE & C\\
Mrk335 & 0.026 & $29.29\pm 0.18$ & $ 75.5\pm  7.7$ & $-1.503\pm 0.077$ & 0 & HST & A
\enddata
\tablecomments{Table~\ref{tab-sampleu} is published in its entirety in the electronic
version of the {\it Astrophysical Journal}. The portion is shown here for guidance 
regarding its form and content.}
\tablenotetext{1}{A value of ``1" indicates that \aox\ for this object is
an upper limit.}
\end{deluxetable}

\begin{deluxetable}{lccccccc}
\tablecolumns{8}
\tabletypesize{\tiny}
\tablecaption{UV and X-ray properties of Sample A objects. \label{tab-sampleax}}
\tablewidth{0pc}
\tablehead{
  \colhead{Object}                          &
  \colhead{$z$}                             &
  \colhead{EW(\feka)}                       &
  \colhead{$\log{l_\nu(2\mbox{\ keV})}$}    &
  \colhead{$\log{L(\mbox{Fe\ K}\alpha})$}   &
  \colhead{$\log{f(\mbox{Fe\ K$\alpha$})}$} &
  \colhead{$\log{L(\mbox{C{\sc iv}})}$}     &
  \colhead{$\log{f(\mbox{C{\sc iv}})}$}    \\ 

  \colhead{}                                &
  \colhead{}                                &
  \colhead{(eV)}                            &
  \colhead{(\luvunit)}                      &
  \colhead{(erg s$^{-1}$)}                  &
  \colhead{(erg s$^{-1}$ cm$^{-2}$)}        &
  \colhead{(erg s$^{-1}$)}                  &
  \colhead{(erg s$^{-1}$ cm$^{-2}$)}        
} \startdata
NGC4593 & 0.009 & $   98.0^{+21.0}_{-21.0}$ & $25.18\pm0.09$ & $ 41.05^{+0.09}_{-0.10}$ & $-12.21^{+0.08}_{-0.11}$ & $41.70\pm0.03$ & $-11.55\pm0.03$\\
NGC3783 & 0.010 & $  120.0^{+14.0}_{-14.0}$ & $25.14\pm0.09$ & $ 41.10^{+0.05}_{-0.05}$ & $-12.98^{+0.12}_{-0.17}$ & $42.50\pm0.01$ & $-10.82\pm0.01$\\
NGC7469 & 0.016 & $  105.0^{+25.0}_{-25.0}$ & $25.28\pm0.09$ & $ 41.18^{+0.09}_{-0.12}$ & $\lesssim -13.19$ & $42.35\pm0.03$ & $-11.43\pm0.03$\\
Mrk1044 & 0.016 & $  186.0^{+61.0}_{-61.0}$ & $24.66\pm0.09$ & $ 40.81^{+0.12}_{-0.17}$ & $-11.84^{+0.22}_{-0.48}$ & $42.02\pm0.01$ & $-11.76\pm0.01$\\
Mrk335 & 0.026 & $\lesssim 54.0$ & $25.38\pm0.09$ & $\lesssim 40.99$ & $\lesssim -14.42$ & $42.75\pm0.02$ & $-11.43\pm0.02$\\
Mrk590 & 0.026 & $  121.0^{+65.0}_{-51.4}$ & $25.25\pm0.09$ & $ 41.21^{+0.19}_{-0.24}$ & $-13.41^{+0.17}_{-0.28}$ & $42.56\pm0.01$ & $-11.64\pm0.01$\\
Mrk290 & 0.030 & $   58.5^{+56.5}_{-43.4}$ & $25.62\pm0.09$ & $ 41.27^{+0.29}_{-0.59}$ & $-12.62^{+0.08}_{-0.09}$ & $42.79\pm0.01$ & $-11.52\pm0.01$\\
Mrk493 & 0.031 & $\lesssim 101.0$ & $25.31\pm0.09$ & $\lesssim 41.19$ & $\lesssim -13.55$ & $42.21\pm0.03$ & $-12.15\pm0.03$
\enddata
\tablecomments{Table~\ref{tab-sampleax} is published in its entirety in the 
electronic version of the {\it Astrophysical Journal}. The portion is shown 
here for guidance regarding its form and content.}
\end{deluxetable}

\begin{deluxetable}{clccrccr}
\tablecolumns{7}
\tabletypesize{\tiny}
\tablecaption{Hypothesis and linear fitting results.\label{tab-fit}}
\tablewidth{0pc}
\tablehead{
  \colhead{Fig}             & 
  \colhead{Sample}          & 
  \colhead{$x$}             &
  \colhead{$y$}             & 
  \colhead{$\rho$\tablenotemark{1}($P_0$\tablenotemark{2})}            &
  \colhead{$k$\tablenotemark{3}}}

\startdata
\ref{fig-civbeffl}   & Combined 
                     & $\log{l_\lambda(2500\mbox{~\AA})}$  
                     & $\log{{\rm EW}_\lambda}$(\civ) 
		     & $-0.559(<0.001)$ 
                     & $-0.198\pm0.015$\\
\ref{fig-civbeffl}   & Combined with $\log{l_\nu(2500\mbox{~\AA})}<30.5$
                     & $\log{l_\lambda(2500\mbox{~\AA})}$  
                     & $\log{{\rm EW}_\lambda}$(\civ) 
		     & $-0.465(<0.001)$ 
                     & $-0.218\pm0.048$\\
\ref{fig-ewciv_aox}  & Combined
                     & \aox 			
		     & $\log{\mbox{EW$_\lambda$(C~{\sc iv})}}$ 
		     & $0.755(<0.001)$ 
		     & $1.035\pm0.075$ \\

\ref{fig-ewfeka_aox} & Sample A
                     & $\alpha_{\rm ox}$ 
		     & $\log{\mbox{EW$_{\rm eV}$(Fe~K$\alpha$)}}$ 
		     & $0.104(0.470)$ 
		     & $0.247\pm0.230$\tablenotemark{4} \\
\enddata
\tablenotetext{1}{Spearman rank correlation coefficient.}
\tablenotetext{2}{Significance levels of Spearman's rank correlation.}
\tablenotetext{3}{Slope from EM algorithm.}
\tablenotetext{4}{Calculated for Sample A using the Buckley-James method in the
ASURV software package \citep{lav92}.}
\end{deluxetable}

\begin{deluxetable}{cccrrr}
\tablecolumns{6} \tabletypesize{\tiny} \tablecaption{Correlation
and partial-correlation analysis results. \label{tab-pcaresult}}
\tablewidth{0pc} \tablehead{
  \colhead{$x$}             & \colhead{$y$}      &
  \colhead{$z$\tablenotemark{a}}             & \colhead{Spearman} &
  \colhead{Pearson}         & \colhead{Kendall}\\
  \cline{1-6}\multicolumn{6}{c}{Combined sample (279)}
}
\startdata
$\log\mbox{\rm EW{(C~{\sc iv}})}$   & $\log{l_\nu(2500\mbox{\ \AA})}$ & \aox 
					 & \nodata
					 & \nodata
					 & $-$0.237\tablenotemark{b}{\ \ \ \ \ \ \ \ \ \ }\\ 
$\log\mbox{\rm EW{(C~{\sc iv}})}$   & $\log{l_\nu(2500\mbox{\ \AA})}$ 
					 &                   
					 & $-$0.599($<$0.001) 
					 & \nodata
					 & $-$0.400($<$0.001) \\ 

\hline

\aox & $\log\mbox{\rm EW{(C~{\sc iv}})}$ & $\log{l_\nu(2500\mbox{\ \AA})}$   
     & \nodata
     & \nodata
     & 0.260{\ \ \ \ \ \ \ \ \ \ }\\ 
\aox & $\log\mbox{\rm EW{(C~{\sc iv}})}$ 
     &            
     & 0.607($<$0.001) 
     & \nodata
     & \nodata\\
\hline

\cutinhead{Combined Sample Without Censored Data (258)}
$\log\mbox{\rm EW{(C~{\sc iv}})}$ & $\log{l_\nu(2500\mbox{\ \AA})}$ & \aox 
				  & $-0.224$
				  & $-0.311$
				  & $-0.220$\\ 
$\log\mbox{\rm EW{(C~{\sc iv}})}$ & $\log{l_\nu(2500\mbox{\ \AA})}$ 
				  &                   
				  & $-0.580$($<0.001$) 
				  & $-0.605$
				  & $-0.417$($<0.001$)\\
\hline
\aox & $\log\mbox{\rm EW{(C~{\sc iv}})}$ & $\log{l_\nu(2500\mbox{\ \AA})}$   
     & $0.332$ 
     & $0.258$ 
     & $0.284$\\ 
\aox & $\log\mbox{\rm EW{(C~{\sc iv}})}$ 
     &            
     & $0.615$($<0.001$) 
     & $0.587$ 
     & $0.450$\\
\hline
\cutinhead{Sample A (49)}

$\log\mbox{\rm EW{(C~{\sc iv})}}$ & $\log{l_\nu(2500\mbox{\ \AA})}$ & \aox 
				  & 0.120(\ \ 0.291)  
				  & 0.074(\ \ 0.214) 
				  & 0.036{\ \ \ \ \ \ \ \ \ \ }\\
$\log\mbox{\rm EW{(C~{\sc iv}})}$   & $\log{l_\nu(2500\mbox{\ \AA})}$ &                   
					 & $-$0.304(0.001) 
					 & $-$0.304(\ \ 0.001) 
					 & $-$0.180(\ \ 0.001)\\

\hline

$\log\mbox{\rm EW{(Fe~K$\alpha$)}}$   & $\log{l_\nu(2\mbox{\ keV})}$ & \aox 
				      & \nodata\tablenotemark{c} 
				      & \nodata\tablenotemark{d} 
				      & $<0.001${\ \ \ \ \ \ \ \ \ \ }\\
$\log\mbox{\rm EW{(Fe~K$\alpha$)}}$   & $\log{l_\nu(2\mbox{\ keV})}$ &                   
				      & $-$0.230(\ \ 0.111)  
				      & \nodata 
				      & \nodata\\

\hline

\aox & $\log\mbox{\rm EW{(Fe~K$\alpha$})}$ & $\log{l_\nu(2\mbox{\ keV})}$   
					   & \nodata 
					   & \nodata
					   & $-$0.130{\ \ \ \ \ \ \ \ \ \ }\\
\aox & $\log\mbox{\rm EW{(Fe~K$\alpha$})}$ &
					   & 0.104(\ \ 0.470) 
					   & \nodata 
					   & \nodata\\

\enddata
\tablecomments{The number in the parentheses after a correlation
coefficient is its significance level. Because we are considering
the null hypothesis,
a small number indicates a possibility of a strong correlation.}
\tablenotetext{a}{$z$ is the controlled parameter. If a $z$ entry is
not empty, we calculate the partial correlation of $x$ and $y$
while controlling for $z$; otherwise, we only calculate the
correlation of $x$ and $y$.} \tablenotetext{b}{For the Kendall's
partial $\tau$ correlation coefficients, generally, the sampling
distribution is unknown; therefore, the probability values are not
available \citep{ken38,ken70}.} \tablenotetext{c,d}{Because EW(\feka) is a set
of censored data, we have to use a generalized correlation
statistics that can deal with censored data to derive the
coefficients. For the non-partial correlations, we used the ASURV
software package \citep{lav92} which only provides Spearman's
$\rho$ and Kendall's $\tau$ correlation correlation coefficients.
For the PCA, we only use the Kendall's partial $\tau$ correlation
statistics \citep{akr96}.}
\end{deluxetable}

\begin{deluxetable}{rcc}
\tablecolumns{3} \tabletypesize{\scriptsize} \tablecaption{The RMS
values of residuals after regression
from different variables. \label{tab-rmsew}} \tablewidth{0pc}
\tablehead{
  \colhead{Independent variables} &
  \colhead{Dependent variable}    &
  \colhead{RMS values}
} 
\startdata 
\luv        & EW(\civ) &0.231\\ 
\aox        & EW(\civ) & 0.228\\
\luv$+$\aox & EW(\civ) & 0.217\\
\hline
EW(\civ)        & \luv\ &0.747\\
\aox            &  \luv\ &0.645\\
EW(\civ)$+$\aox &  \luv\ &0.615
\enddata
\tablecomments{To consistently compare the RMS values, we compute them using
the combined sample without the censored data.}
\end{deluxetable}

\begin{deluxetable}{ccc}
\tablecolumns{3}
\tabletypesize{\scriptsize}
\tablecaption{Correlation and regression analysis for EW, emission line
luminosity, and flux data between \feka\ and \civ.
\label{tab-cfeciv}}
\tablewidth{0pc}
\tablehead{
  \colhead{Relations}  &
  \colhead{$\rho$($P$)\tablenotemark{1}} &
  \colhead{$k$\tablenotemark{2}}
}
\startdata
EW        & 0.319(\ \ 0.027) & 0.291$\pm$0.131\\ 
$L$(line) & 0.529($<$0.001)  & 0.588$\pm$0.079\\ 
$f$(line) & 0.551($<$0.001)     & 0.978$\pm$0.188\\ 
\enddata
\tablenotetext{1}{Correlations are tested using Spearman's $\rho$ and the
significance level ($P$) is evaluated against the null hypothesis.}
\tablenotetext{2}{We use the Buckley-James method to do linear
regression and $k$ is
the slope. The computation is done using the ASURV software
package \citep{lav92}}
\end{deluxetable}